\documentclass[fleqn,10pt]{wlscirep}
\usepackage[utf8]{inputenc}
\usepackage[T1]{fontenc}
\usepackage[graphicx]{realboxes}
\usepackage{subcaption}
\usepackage{times}
\usepackage{amsmath}
\usepackage{color}
\usepackage{overpic}
\usepackage{textcomp}

\title{Assessing the interplay between human mobility and mosquito borne diseases in urban environments}

\author[1*]{Emanuele Massaro}
\author[2]{Daniel Kondor}
\author[2,3]{Carlo Ratti}
\affil[1]{HERUS Lab, École Polytechnique Fédérale de Lausanne, Lausanne (CH)}
\affil[2]{Singapore-MIT Alliance for Research and Technology, Singapore}
\affil[3]{MIT Senseable City Lab, Cambridge (MA), USA}

\affil[*]{emanuele.massaro@epfl.ch}

\begin{abstract}
Urbanization drives the epidemiology of infectious diseases to many threats and new challenges. In this research, we study the interplay between human mobility and dengue outbreaks in the complex urban environment of the city-state of Singapore. We integrate both stylized and mobile phone data-driven mobility patterns in an agent-based transmission model in which humans and mosquitoes are represented as agents that go through the epidemic states of dengue. We monitor with numerical simulations the system-level response to the epidemic by comparing our results with the observed cases reported during the 2013 and 2014 outbreaks. Our results show that human mobility is a major factor in the spread of vector-borne diseases such as dengue even on the short scale corresponding to intra-city distances. We finally discuss the advantages and the limits of mobile phone data and potential alternatives for assessing valuable mobility patterns for modeling vector-borne diseases outbreaks in cities.
\end{abstract}
\begin{document}

\flushbottom
\maketitle

\thispagestyle{empty}

\section*{Introduction}
\label{Sec:intro}

Rapid urbanization and increased mobility brings new challenges for epidemics~\cite{neiderud2015urbanization}. Estimates show that more than half of the world's population already lives in cities, while further big increases are expected especially in Asia and Africa. Challenges presented by new megacities include the rapid spread of new epidemics, which can become worldwide threats due to increased global connectivity~\cite{Brockmann2013, gvr2017, massaro2018resilience}. Poor housing conditions in rapidly growing cities can exacerbate epidemic threats, especially in the case of insect and rodent vector diseases and geohelminths~\cite{world2010hidden, ajelli2017modeling}. Governments need to look for innovative solutions for monitoring and controlling epidemics~\cite{lindsay2017improving}. An important part of these considerations is understanding the relationship between disease spread and human mobility, which have been previously linked on global scales~\cite{Brockmann2013, gvr2017}. In this paper we explore the effectiveness of pervasive technologies, specifically mobile phone data, in predicting and understating the emergence of mosquito borne disease outbreaks in urban environments. Cell phone data has been shown to be valuable in monitoring mobility patterns in near real-time~\cite{kung2014exploring}. Such information has a large potential in epidemiological modeling and control~\cite{wesolowski2014quantifying}, yet it has been often unreliable and difficult to obtain with traditional methods, especially in developing countries with rapidly changing urban environments and limited resources to conduct travel surveys.

We study the influence of human mobility on the spread of the mosquito-borne dengue virus, as inferred from a large-scale mobile communication dataset in the city-state of Singapore. Contrary to previous studies that either focused on this problem at the scale of countries or regions~\cite{cummings2004travelling,lourencco20142012,wesolowski2012quantifying,teurlai2012can,Brockmann2013,wesolowski2014quantifying,WesolowskiPNAS2015}, essentially treating cities as well-mixed nodes in a larger travel network, or used small-scale data of human movement inside cities collected through surveys~\cite{stoddard2013house} or only use theoretical models and aggregate on intra-city human mobility~\cite{stoddard2009role,kong2018modeling,karl2014spatial,zellweger2017socioeconomic,telle2016spread}, we now employ a large-scale dataset of human mobility to study the connection between intra-city mobility and dengue spread. We focus on comparing a dengue transmission model based on people's real commuting patterns (as inferred from the mobile phone dataset) with the observed dengue cases and with simulations employing random mobility models. This allows us to measure the impact of mobility model on the accuracy of modeling the spatial distribution of dengue cases. We especially focus on comparing random mobility that results in perfect mixing of population with more structured mobility models, effectively evaluating the importance of intra-city human mobility in dengue spread.

Dengue fever is a mosquito-borne viral infection, transmitted by female mosquitoes of the species Aedes aegypti and Aedes albopictus when biting humans. The infection causes flu-like symptoms with occasional complications that can be fatal. There are four strains of the virus and the infection with one strain produces lifelong immunity to that type. However, a second infection with a different type increases the risk of severe complications. Dengue continues to be a global threat, with about half the world's population being at risk of infection~\cite{guzman2010dengue}. Worldwide, there are more than 50 million infections every year, leading to half a million hospitalizations and up to 25 thousands deaths. Dengue is prevalent in tropical and sub-tropical climates worldwide, mostly in urban and semi-urban areas. The prevention and control solely depends on controlling the mosquito populations. 
There is active development for vaccines, with a first-generation vaccine becoming available recently~\cite{rothman2016dengue}. Dengue affects Singapore in particular and two major outbreaks were observed in 2013 and 2014 (\figurename~\ref{fig:tempCases}(a)).

The modeling of dengue outbreaks has attracted the interest of many researchers in many disciplines from physics to computational biology. Presented models investigate, for example, the variability of the mosquito population~\cite{esteva1998analysis}, the variability of the human population~\cite{esteva1999model}, the vertical transmission between mosquitos (that is, the transmission between mosquito generations)~\cite{esteva2000influence} as well as seasonal patterns~\cite{hartley2002seasonal}. Otero at al.~\cite{otero2006stochastic, otero2008stochastic} presented a dengue model, which takes into account the evolution of the mosquito population. Another study shows that dengue appears to travel in waves~\cite{cummings2004travelling}. As the flight range of mosquitoes is limited to a few hundred meters~\cite{stoddard2013house}, it is generally assumed that humans carry the dengue virus to previously dengue-free areas and infect local mosquitoes. There is evidence that the spread of mosquito-borne diseases is related to human mobility~\cite{stoddard2009role}. Various agent-based simulations suggest that the mobility of humans could be the main driving force behind the spread of the dengue virus~\cite{barmak2011dengue,de2011modeling}. Teurlai et al.~\cite{teurlai2012can}  showed that the human mobility, estimated from the road network, influences the spread at a national scale in Cambodia. Especially house-to-house human movements seem to play a key role in Iquitos, Peru~\cite{stoddard2009role}. Related malaria studies show that human mobility, which is estimated from cell-phone networks, drives the dissemination of malaria parasites as well~\cite{wesolowski2012quantifying}. Recently Wesolowski et al. studied the impact of human mobility on the emergence of dengue epidemics in Pakistan~\cite{WesolowskiPNAS2015} using mobile phone-based mobility. 

Considering the threat presented by dengue especially in cities, many authors studied the effect of dengue fever in urban environments~\cite{kong2018modeling, karl2014spatial, lourencco20142012, zellweger2017socioeconomic, telle2016spread, lindsay2017improving}. While these work generally assume that intra-city mobility is an important factor for dengue epidemics, a direct quantification of this effect is still lacking. For the first time, we analyze the effectiveness of mobile phone useage data to predict the dengue spreading in an urban environment, such as Singapore. In doing so, we compare random mobility patterns with the real one estimated from anonymized mobile phone usage records in an agent-based model of dengue transmission adapted from previous studies~\cite{lourencco20142012,WesolowskiPNAS2015}. This way, we are able to characterize the effect of human mobility on urban scales in the spread of vector-borne diseases and the effectiveness to use mobile phone data to estimate disease epidemics on this scale as well.

\section*{Results}
\label{Sec:res}

\begin{figure}[t!]
    \centering
    \includegraphics[width=1\linewidth]{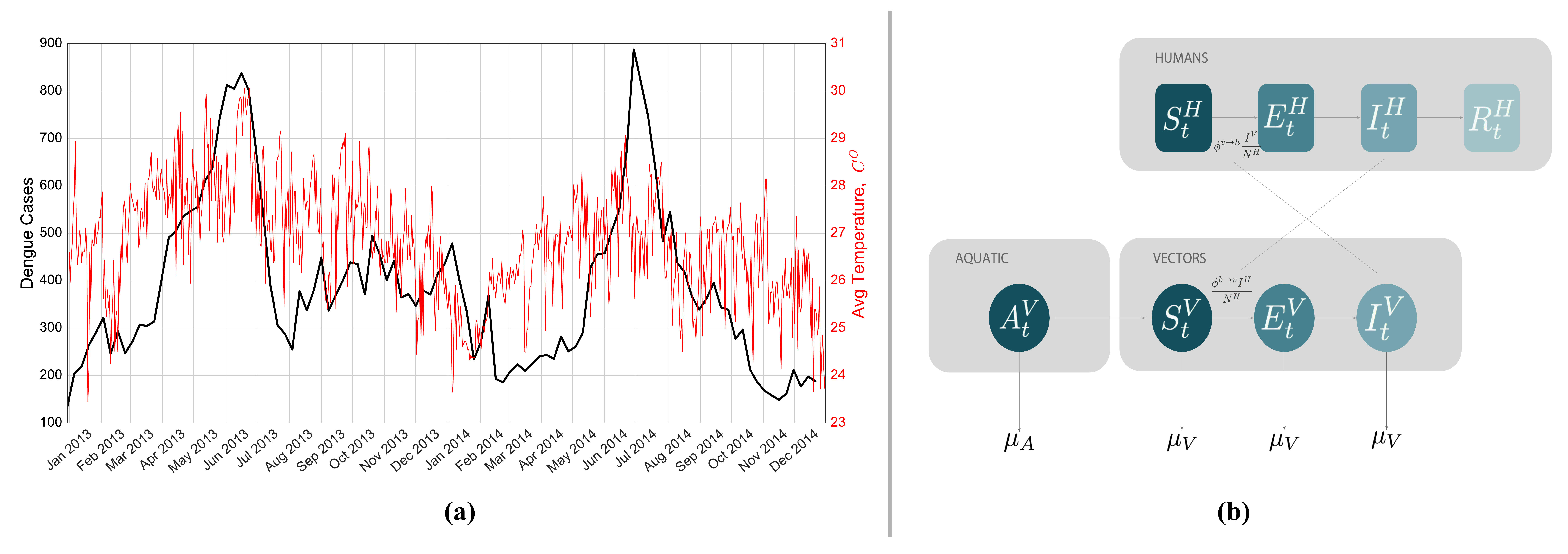}     
\caption{\textbf{Temperature Dependency of the Dengue cases and Schematic representation of the Human-Vectors interactions in the epidemiological model}. (a) Weekly observed dengue cases and average temperature in Singapore from January 2013 to December 2014. Two outbreaks took place during those two years during the summer. It is possible to observe a correlation between temperature and number of reported cases of people affected by the disease. (b) Compartmental classification for DENGUE disease. Humans can occupy one the four top compartments: susceptible, which can acquire the infection through contacts (bites) with infectious mosquitoes; exposed, where individuals are infected but are not able yet to transmit the virus; infectious, where individuals are infected and can transmit the disease to susceptible mosquitoes; and recovered or removed, where individuals are no longer infectious. The density of mosquitoes changes according to the seasonal transition from Aquatic (A) to Adult Mosquitoes (V). Similar to the humans case, Mosquitoes can occupy three different compartments and they can die with a given rate depending on the temperature. }
\label{fig:tempCases}  
\end{figure}

\begin{figure}[b!]
    \centering  
    \includegraphics[width=1\linewidth]{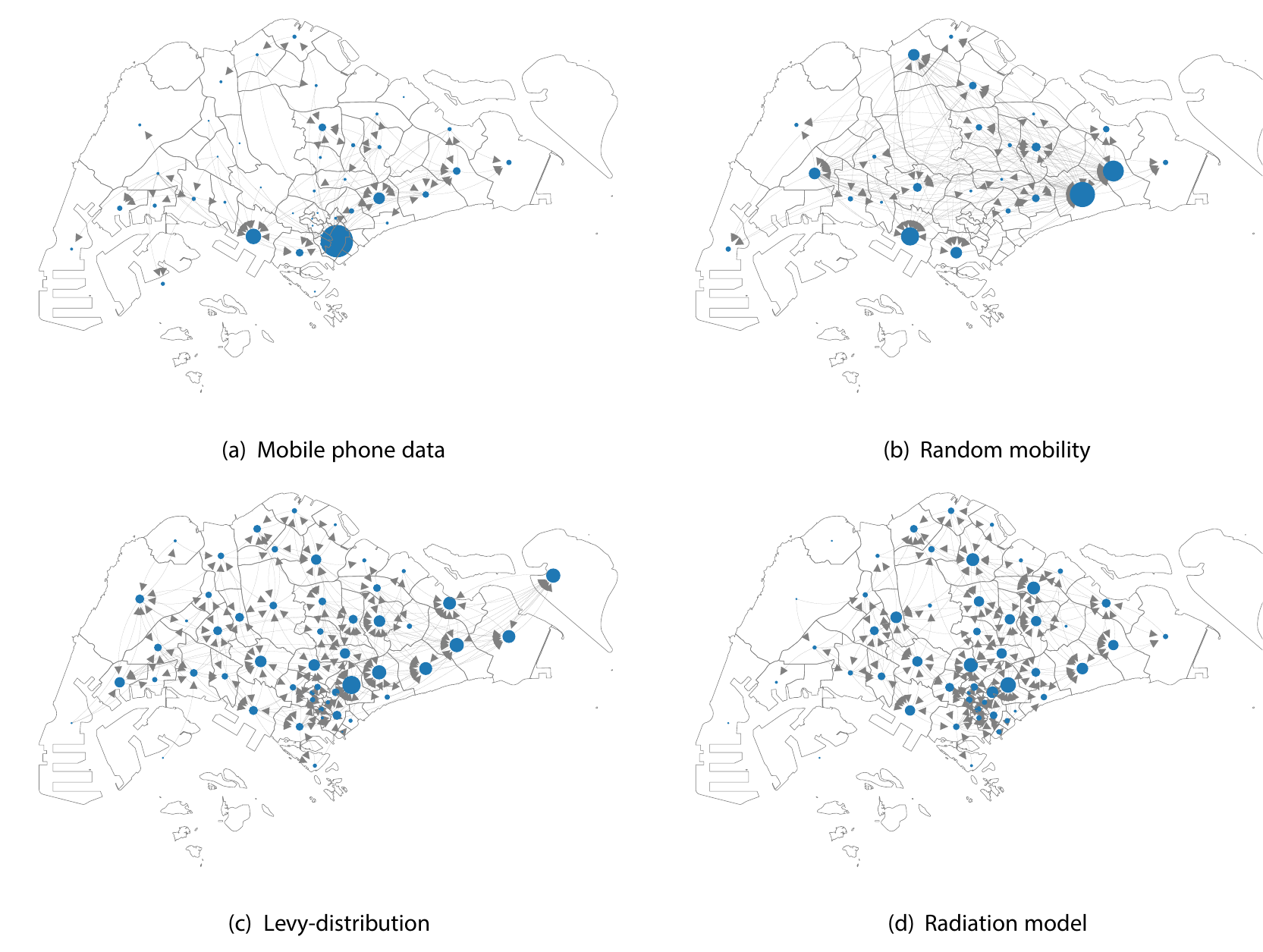} 
\caption{\textbf{Commuting flows from home to work locations aggregated at the 55 planning areas}. The location of the nodes corresponds to the centroid of the areas and their size corresponds to the incoming degree which corresponds the total amount of agents that commutes everyday to that area. In this figure we report only the most significant nodes in terms of incoming flow (i.e. greater than 95th percentile the distribution). (a) We can observe that major hub in the mobile phone data mobility model corresponds to the  Central Business District where the majority of the jobs are located. (b) The random mobility  mobility has different hubs randomly distributed  in the space. (c) The Levy-distribution  and (d) the radiation model show similar patterns, with an homogeneous distribution on the territory without significant hubs: however the mobility derived from the radiation model is more aggregated in the central part of the city.}
\label{fig:flowsMap}  
\end{figure}

We propose an agent-based dengue transmission model in which humans and mosquitoes are represented as agents and humans go through the epidemic states of dengue~\cite{esteva1998analysis,esteva1999model,esteva2000influence,barmak2011dengue}. 
To model dengue dynamics, we use a stochastic population model based on the ordinary differential equation (ODE) framework employed by Lourenco and Recker to describe a dengue outbreak in Madeira, Portugal~\cite{lourencco20142012} and then used by Wesolowski and colleagues to model the dengue outbreak in Pakistan~\cite{WesolowskiPNAS2015}. The epidemiological model depends on both temperature-dependent and constant parameters as described in the Methods section and reported in~Table~\ref{table:tempPar} and~Table~\ref{table:constPar}. We employ an agent-based approach for humans, while we model localized mosquito subpopulations stochastically. As a necessary simplification, we only consider one serotype of dengue; in this case, individuals can only be infected once. The physical environment in which the epidemic takes place is a regular grid, composed of $320\,\mathrm{m}\times 320\,\mathrm{m}$ cells, overlaid the city of Singapore.

The model is composed of two phases: (i) the phase  of \emph{reaction}, defined by the epidemioloigcal model (see the section Materials and Methods for details and \figurename~\ref{fig:tempCases}(b) for a schematic overview), where disease transmission takes place in each grid cell; (ii) the phase of \emph{diffusion} where agents are moved from one grid cell to another according to the mobility model under consideration: the mobility flows aggregated at census district level for the different mobility models are reported in~\figurename~\ref{fig:flowsMap}. Each day consists of two reaction phases, corresponding to day and night, and two diffusion phases, corresponding to people's morning and evening commute.

In this work, we consider four different mobility models (see~\figurename~\ref{fig:flowsMap}) and compare their predictive power about the dengue outbreaks of 2013 and 2014 in Singapore. In each mobility model, we assign a home and work location (grid cell) to each agent who are assumed to commute between these two daily. The models differ in the way how this assignment is made: (1) mobile phone data: we use anonymized call detail records of one mobile phone operator in Singapore, collected in a two month period in 2011 that allows us to estimate home and work cells for $2.3$~million agents; (2) random work location: in this case, we still use the home cells estimated from the mobile phone data, but work locations are assigned randomly; (3) Levy-distribution: each agent is assigned a random home location based on the mobile phone data and a work cell is chosen such that the commuting distance follows a truncated Levy-distribution; (4) radiation model: we use census data to distribute the home locations of agents~\cite{singapore2010census} and then we choose work cell locations according to the radiation model of Simini~et~al.~\cite{simini2012universal}. In total, there are 2,598 grid cells with either a home or work location in them. More detailed description of the mobility models is given in the Materials and Methods section, while we present a comparion between the mobility models in the Supplementary Material, in Figure S1 to Figure S5.. Most notably, flows of people on the district level are highly correlated among the mobile phone data and the radiation model ($r = 0.938$), somewhat less correlated with the Levy-distribution model ($r = 0.901$) and significantly less correlated among mobile phone data and random mobility ($r = 0.304$). This way, we conclude that the radiation, Levy-distribution and random mobility models give successively worse approximations of real mobility patters.

Beside the mobility model, we have two main variable parameters, the number of mosquitoes per human, $x_v$ and average bite rate of mosquitoes, $a$ (more thorough definitions of these and a discussion on model parameters are given in the Materials and Methods section). We perform a sensitivity analysis on these, by exploring the phase space $x_v \in [0.004, 0.1]$ and $a \in [0.14, 0.26]$. This allows us to calibrate our model to the population of agents in Singapore; this is a necessary step since exact estimation of these parameters is especially challenging in real-world condition, while several parameters in the epidemiological model cannot be reliably measured in real-world conditions, only in controlled laboratory experiments~\cite{liu2014vectorial}. In our approach, we use best available estimates from the literature for most parameters, while allow variation of $x_v$ and $a$ to deal with this inherent uncertainty. We select the best parameter combination for each mobility model to evaluate our results.

We start our simulations with initial conditions for infected human agents based on the observed number and distribution of cases in January 2013, while we obtain the initial mosquito populations by running the population dynamic model for an initial warm-up period as described in the Materials and Methods section. To account for the stochastic nature of the simulation, for each parameter value, we ran the simulation 100 times and report the median and average values in the following.

\subsection*{Temporal Analysis}
\begin{figure}[t!]
    \centering
    \includegraphics[width=1\linewidth]{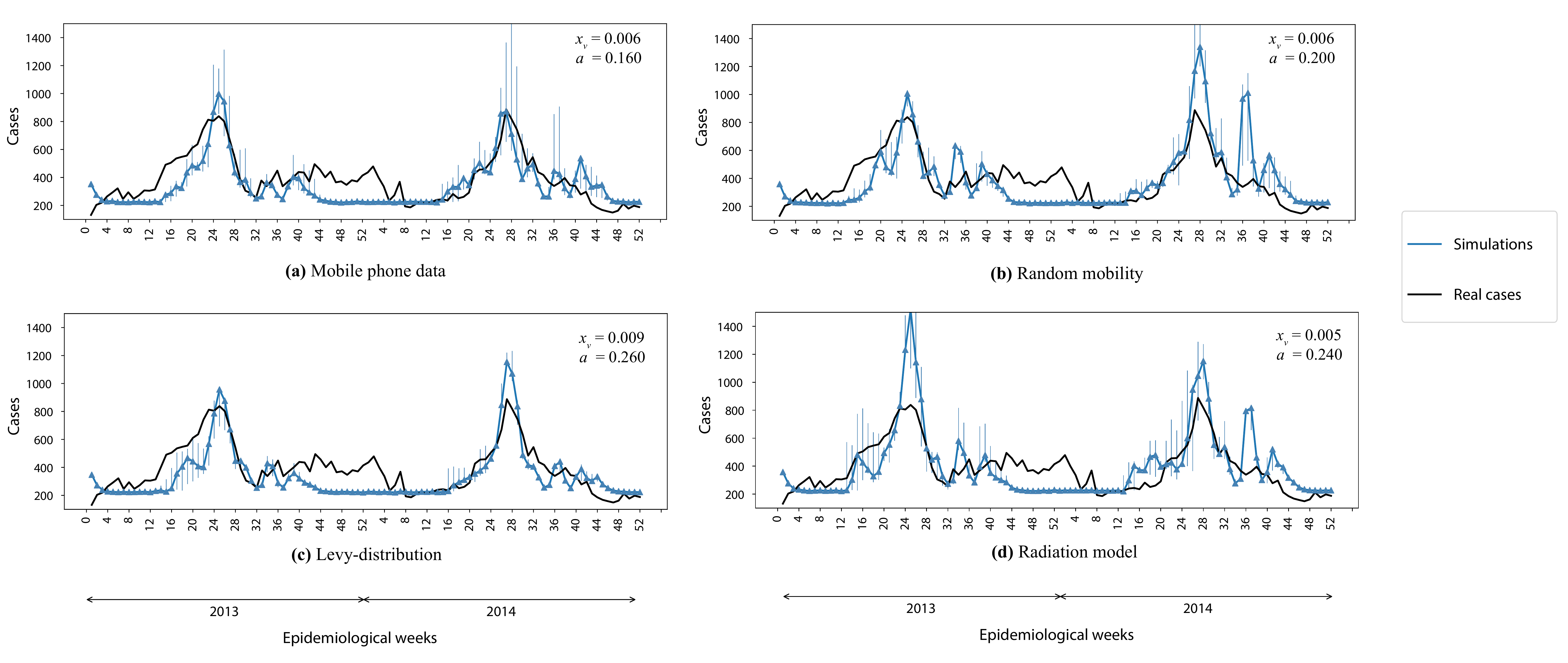}
\caption{\textbf{Temporal analysis}. We report the comparison between the best simulated scenario and the observed number of dengue cases during the 2013-2014 outbreaks. Parameter values for $x_v$ (average number of mosquitoes per human) and $a$ (mosquito bite rate) are displayed in the figure legends for each case.}
\label{fig:simCases2013}  
\end{figure}

\begin{table}[h!]
\centering
\caption{Prediction error $R_2$ for the best couple of the parameters $x_v$, $a$ for the different models. }
\label{tab:rmodels}
\begin{tabular}{l l l l}
Mobility Model  & $R_2$ & $x_v$ & $a$ \\
\hline \\
Mobile Phone & 0.65 & 0.006 & 0.16 \\
Random & 0.51 & 0.006 & 0.2 \\
Levy Distribution & 0.62 & 0.009 & 0.26 \\
Radial Model & 0.56 & 0.005 & 0.24 \\
\hline
\end{tabular}
\end{table}
In~\figurename~\ref{fig:simCases2013} we report the comparison of the number of observed cases and the median of the simulated infected cases estimated from our simulations during the epidemiological weeks in 2013 and 2014 for the four different mobility models. In particular for each mobility model we report the pair of parameters $x_v$-$a$ that maximize the $R^2$ between the simulations and observed number of cases. Each mobility model is able to predict quite well the temporal evolution of the dengue outbreaks, since the epidemiological dynamics mainly depends on the value of the temperature. Each mobility model optimizes the prediction for different values of the parameter  $x_v$-$a$ as reported in the legends of~\figurename~\ref{fig:simCases2013}. In order to find the best pair of parameter values, we compute the $R^2$ between the observed and the predicted number of cases between the 12th and 26th epidemiological weeks when the epidemic peaked during the study period. We show optimal $R^2$ values in~Table~\ref{tab:rmodels} and display variation of $\log R^2$ in the phase space in Figure S13 in the Supplementary Information. The Mobile Phone Data and the Levy Distribution mobility models have the better accuracy with a value of $R^2$ of $0.65$ and $0.61$ respectively while the Random and the Radiation mobility models tend to overestimate the number of cases and with  $R^2$ of $0.52$ and $0.57$ respectively. Nevertheless, we still conclude that all models reproduce the main trends in the epidemic well.

\subsection*{Spatial Analysis}
\begin{figure}[h!]
    \centering
    \includegraphics[width=.5\textwidth]{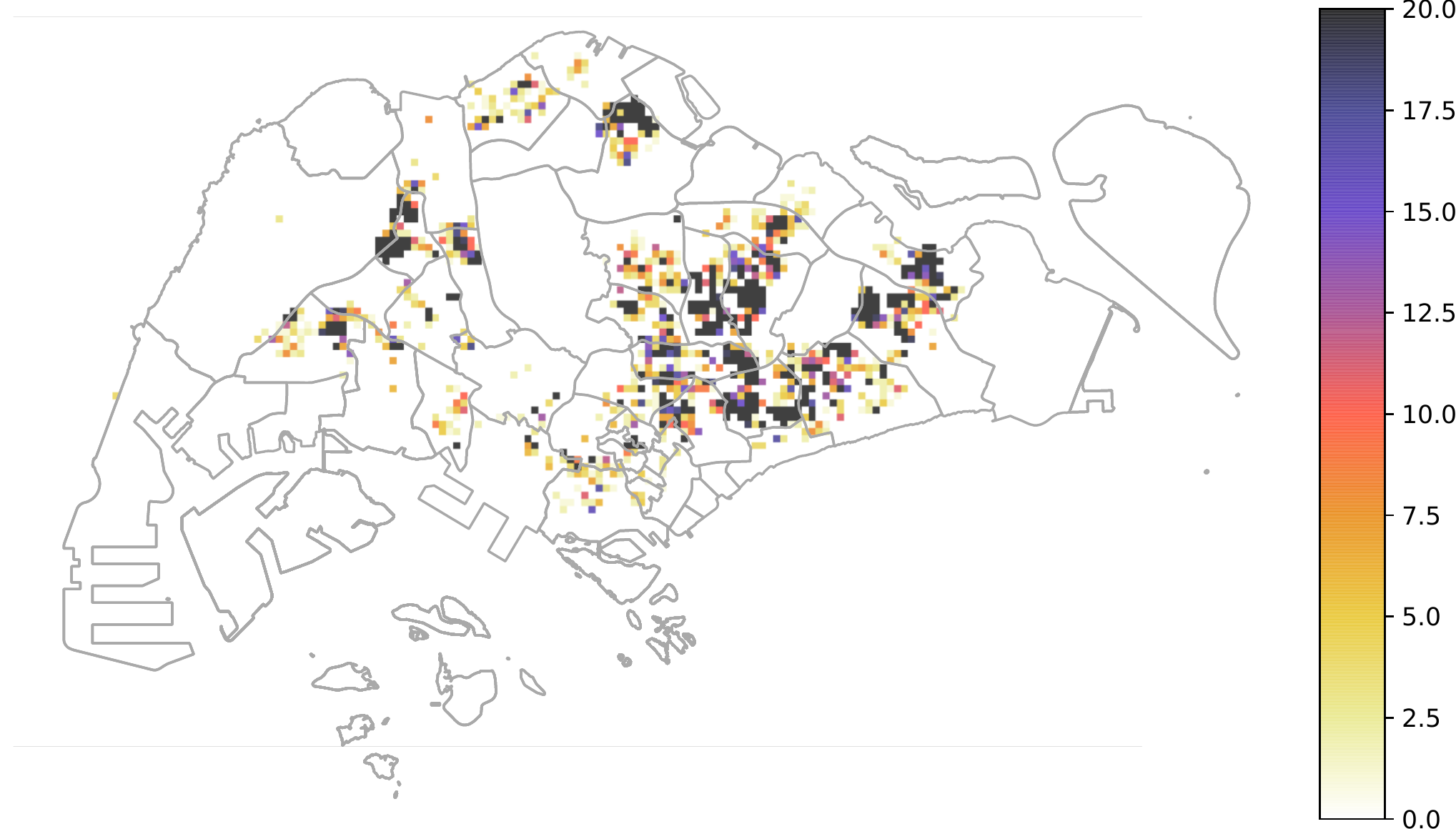}
    \caption{\textbf{Observed dengue cases}. Cumulative spatial distribution of observed dengue cases during the 2013 and 2014 outbreaks.}
    \label{fig:spatCaseReal}
\end{figure}

\begin{figure}[h!]
    \centering
    \includegraphics[width=1\textwidth]{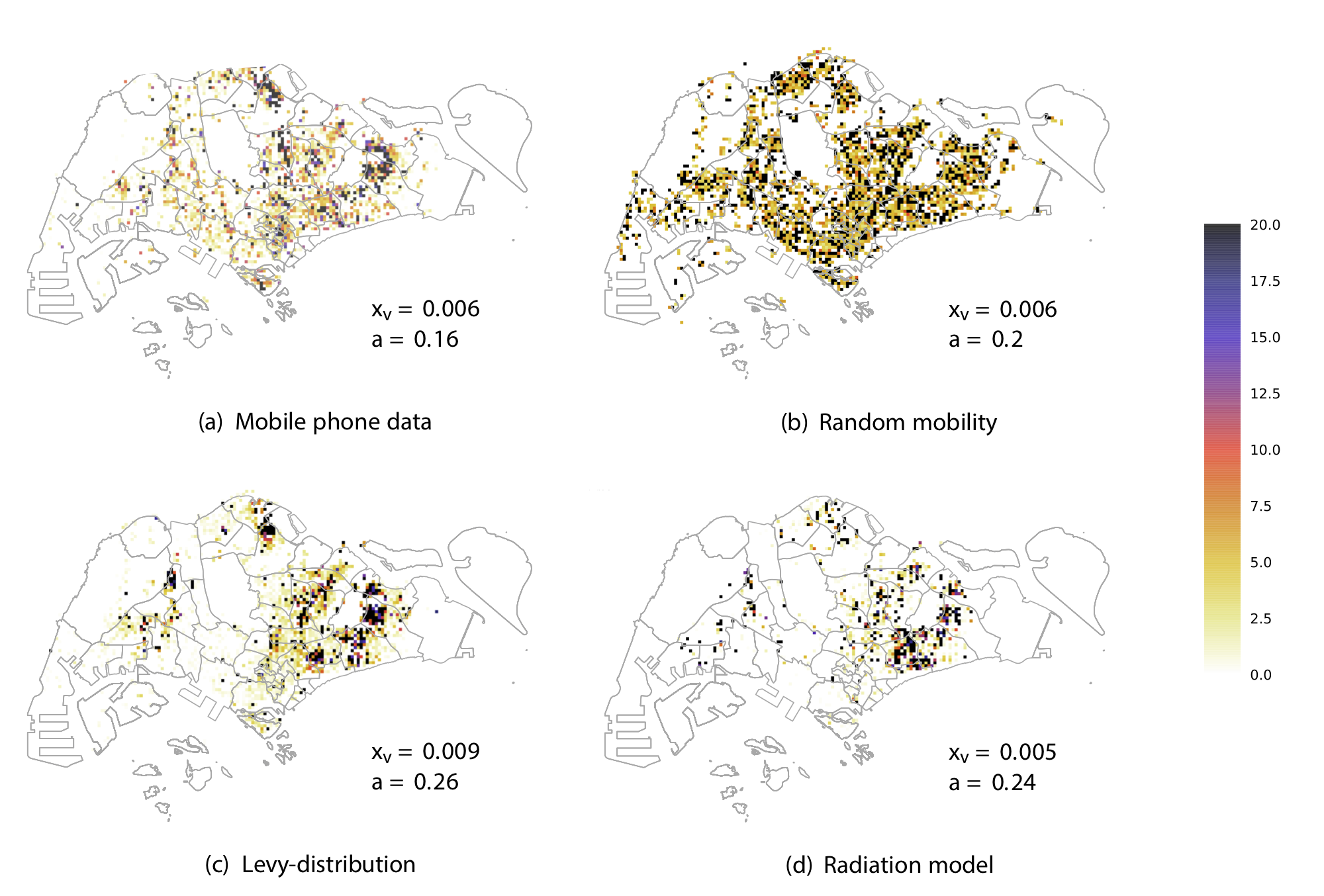}
\caption{\textbf{Spatial analysis}. We report the heatmap of the cumulative number of cases for the four mobility models. For each simulated scenario we report the results with the best parameter values, as shown in the figures.}
\label{fig:spatCase}  
\end{figure}

In this section we show the results of our simulations and we compare it with the spatial distribution of number of reported cases in 2013-2014 in Singapore. The distribution of \emph{Ae. aegypti} expanded during the decade from 2003 to 2013 and the percentage of houses with mosquito breeding in 2013 and 2014 was significantly higher than in previous years~\cite{hapuarachchi2016epidemic}. As expected, the dengue case distribution pattern in 2013 and 2014 was in line with the geographical spread of \emph{A. aegypti} in the country~\cite{hapuarachchi2016epidemic}. The biggest clusters remain in Tampines in the eastern part of the island, however more are now in the west and north. In order to quantify the effect of human mobility on the spatial propagation the dengue in Singapore, we compare the results of our model with observed case scenarios by considering the four different mobility patterns: (1) mobile phone data; (2) random; (3) Levy; (4) radiation. We show the cumulative spatial distribution of observed cases in \figurename~\ref{fig:spatCaseReal} and simulated cases in the four models in \figurename~\ref{fig:spatCase} with with the $x_v$ and $a$ parameters that give the best estimate for the temporal patterns (as reported in~\figurename~\ref{fig:simCases2013}).
We can see that the mobility  plays an important role for predicting the spatial distribution of the number of cases. Indeed the spatial distribution of the number of cases predicted by the random mobility model is uniformly distributed among the city, while the other mobility models allow us to detect key hotspots of the outbreaks similar to the observed scenario.

\begin{figure}[t!]
\centering
\includegraphics[width=0.75\linewidth]{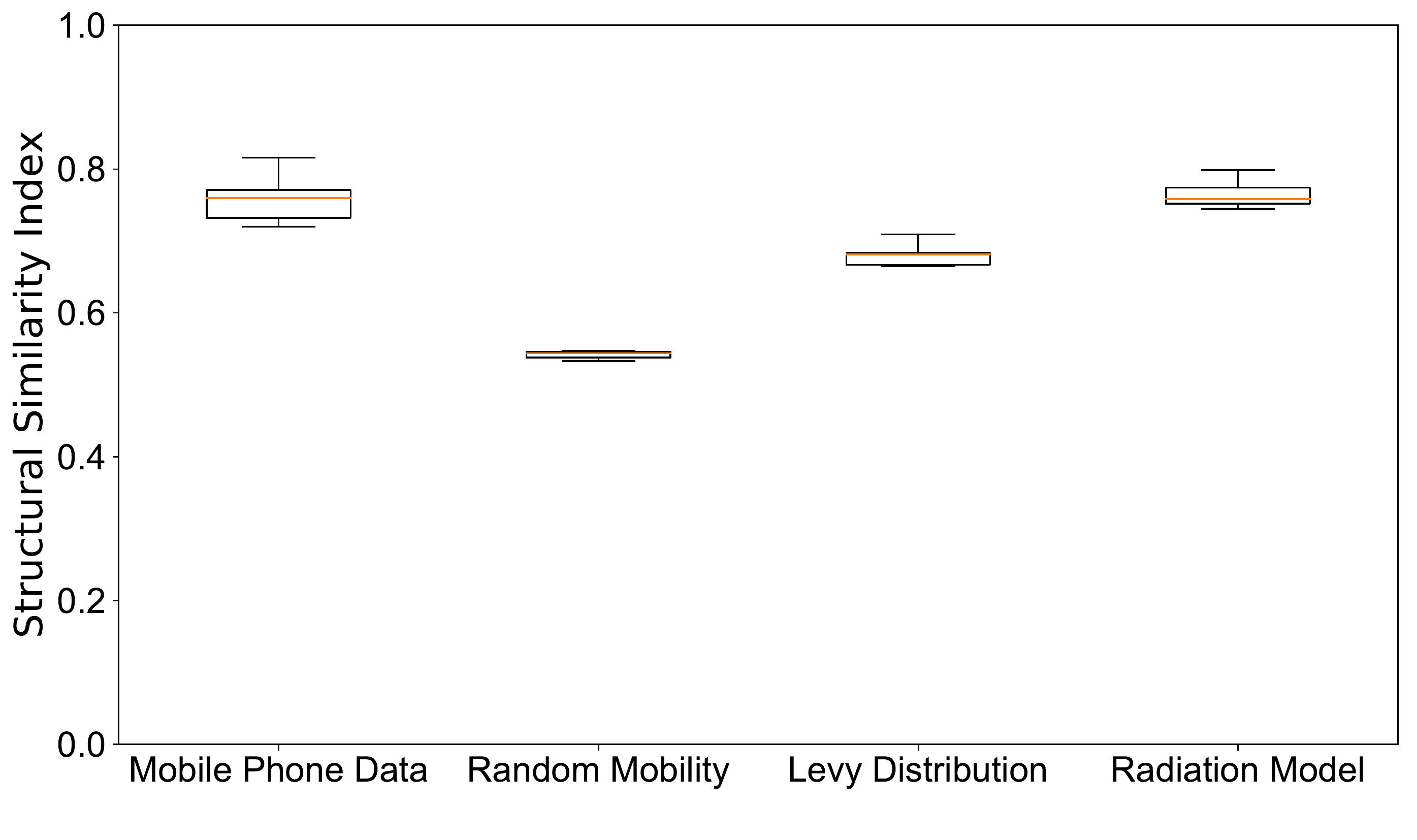}
\caption{\textbf{Spatial analysis}. Boxplot of the value of the SSIM Index for each weeks during the  2013-2014 outbreaks using the best parameter $x_v$-$a$ shown in~\figurename~\ref{fig:simCases2013}. SSIM index values were calculated for each epidemiological week during the outbreak for each of the 100 simulation runs. The distributions of these values are shown as boxplots for each mobility model in this figure. The boxplots show the minimum, first quartile, median, third quartile and maximum among the SSIM values observed. We see that in all cases, the range of data is quite small; the mobile phone data and radiation model results are clearly distinguished from the random mobility and Levy-distribution results.}
\label{fig:ssim}  
\end{figure}

To better distinguish between the predictive power of different mobility models, we computed structural similariy (SSIM) scores~\cite{Wang2004,wang2009mean} (see the Supplementary Information for a description) for each case in each epidemiological weeks, with the best parameters $x_v$-$a$ and compare their distribution in~\figurename~\ref{fig:ssim}. We can observe that the mobile phone mobility model and the radiation model perform in a similar way, consistently well approximating the observed spatial distribution during the time period of our study. The Levy-distribution model is performing slightly worse, while the random mobility model gives significantly worse results. Looking at the results in~\figurename~\ref{fig:spatCase}, we find that the overall distribution of the infected cases for the random mobility model corresponds well to the average population density (i.e.~average of work and home locations in each cell). This means that our random mobility model achieves a good mixing among the population. The difference from the real distribution of dengue cases and the other mobility models highlights that uniform mixing among the population does not account for a spread of dengue in Singapore, thus mobility patterns are an important factor. While previous large-scale epidemiological studies often treat cities as well-mixed nodes in a larger travel network~\cite{cummings2004travelling,lourencco20142012,wesolowski2012quantifying,teurlai2012can,Brockmann2013,wesolowski2014quantifying,WesolowskiPNAS2015}, our results show that disease spreading can exhibit important localized patterns inside cities as well, in line with studies done previously on smaller samples of the population or aggregate models of human mobility~\cite{stoddard2009role,stoddard2013house,kong2018modeling,karl2014spatial,zellweger2017socioeconomic,telle2016spread}. It is unclear yet, how the intra-city and inter-city epidemiological models are best reconcilied; we note that frequent travelers are often a non-uniform sample of the total population of any city, thus local and long-range spread of infectious diseases can have complex intervowen patterns. The further difference between the Levy and radiation mobility model is consistent with previous work which found the radiation model to best reproduce the statistical properties of human commuting~\cite{simini2012universal}. Furthermore, the good results obtained from the mobile phone data show that the home-work commuting estimated from this dataset indeed accounts for the most important factors in human mobility in Singapore.

\section*{Discussion}
\label{Sec: disc}
More than $80\%$ of the world’s population is at risk from at least one vector-borne disease~\cite{gvr2017}. The populations most at risk are those living in poverty in the tropical and subtropical areas, but as the case of Singapore shows, highly developed cities and countries still need continued efforts to prevent outbreaks~\cite{hapuarachchi2016epidemic}. The rapid urbanization, the increase in international travel and trade, the modification of agriculture and environmental changes have increased the spread of vector populations, putting more and more people at risk. Mobile phone data can give real-time mobility information that can be combined with infectious disease surveillance data and seasonally varying environmental data to map these changing patterns of vulnerability in cities that are changing everyday. In this paper we proposed an agent based model in order to explicitly simulate the epidemic spread of the disease as governed by the transmission dynamics of the dengue virus through human-mosquito interactions and promoted by the population movements across the city state of Singapore. In this methodology humans and mosquitoes are represented as agents and humans go through the epidemic states of dengue.

We modelled four different mobility patterns: 1) mobility estimated from mobile phone data, 2) random mobility patterns, 3) mobility estimated from census data following a Levy distribution model and 4) mobility estimated from census data following a radiation model. We were able to reproduce the main temporal and spatial patterns of the dengue outbreak in 2013 and 2014. Our results show that human mobility is a very important factor in the spread of vector-borne diseases such as dengue even on the short scale corresponding to intra-city distances. This is evidenced by the large difference found between the observed spatial pattern of dengue cases and the ones obtained by the completely random mobility model which corresponds to a ``perfect mixing'' among the population. This extends the results obtained from the previous work of Wesolowski~et~al.~\cite{WesolowskiPNAS2015} who showed how human mobility determines the spread of dengue on the scale of a country and studies that investigated the relationship between human mobility and spread of vector diseases on different spatial scales~\cite{kong2018modeling,wesolowski2014quantifying,wesolowski2012quantifying, lindsay2017improving, telle2016spread, zellweger2017socioeconomic, maneerat2017agent}. We believe that our main contribution is showing that human mobility patterns are important for the spread of vector-borne diseases even on intra-city scales; this is in contrast to previous studies which often assume cities to be a well-mixed environment for the purpose of epidemiology and study disease spreading between cities and regions~\cite{Brockmann2013,wesolowski2014quantifying,lourencco20142012}. It is an interesting question for future work to what extent this result applies to other types of diseases, e.g.~airborne infections that require only shorter co-location of people to spread, thus are able to exploit mixing of population in a more rapid way.

Furthermore, we found that more sophisticated models of intra-city mobility can give good estimates of the spatial spread of dengue, opening up the possiblity to incorporate these into modeling and control of vector diseases in urban environments. The proposed methods could be integrated into urban planning in near real time. Mobile phone data is an obvious candidate for this purpose, giving real-time information on people's movements. A major limitation of mobile phone data generated by national operators is the difficulty in capturing cross-border travel patterns and it is not possible to monitor with high accuracy the flux of people travelling to the city. On the other hand, we found that the radiation model of people's commuting behavior performs similarly well, opening up the possibility to improve prediction of disease spread if accurate census data is available. Thus, we believe our methods can be readily used in other cities where these mobility models can be estimated, while accuracy will be affected by overall predictability of human movements and regularity of commuting patterns~\cite{gonzalez2008understanding,returners,simini2012universal,Song1018,kung2014exploring}. Concluding, we note that the methods we presented here can be readily generalized to consider different mosquito-borne diseases such as dengue, chikungunya, malaria, yellow fever and different sources and models of human mobility, having a large potential usability for better understanding, control and prevention of vector disease epidemics in urbanized areas.

\section*{Materials and Methods}
The code used for our simulations is available online~\cite{CODE}.
\label{Sec: method}
\subsection*{Mobile phone data}

Anonymized call detail records (CDRs) were collected over a two month period in 2011 by one of the mobile phone operators in Singapore with a significant market share (a statistical analysis is reported in Figure S9 in the supplementary materials.). The data includes more than $2$ billion records in total and includes the approximate time and location of events (including phone calls and text messages). Locations are collected at the cell tower level with further noise applied for privacy reasons. We use this data to assign two ``favorite'' locations to each user: (i) home and (ii) work. In Singapore, according to a study by the Land Transport Authority, about $80\%$ of all trips go to either a work or a home location\cite{holleczek2014detecting}. This implies that the infection with the dengue virus in Singapore very likely happens either at home or at work, thus we focus on commuting between these two locations when modeling human mobility in this paper. To estimate home and work locations, we perform a spatial clustering of the CDRs, creating overlapping clusters of events which are spatially close to each other (a threshold of $500\,\mathrm{m}$ was used so as to account for the potential uncertainty regarding which one of nearby antennas a phone connects to). To be able to distinguish between home and work locations, we performed this clustering procedure separately for records generated between 8pm and 6am on weekdays and during weekend (for home locations) and records generated between 10am and 4pm on weekdays (for work locations). After this procedure, we selected the largest clusters for both cases and filtered the list of users who had at least 10 events in both clusters. Following this procedure, there are $2,307,230$ users to whom we can assign a home and a work location. We then assign users' home and work locations into a $320m \times 320m$ grid overlay $G$ which we use as the basis of the epidemic simulation. We display the distribution of these home and work locations in Figure S2, while we show the nonempty grid cells in Figure S12 in the Supplementary Material.

To show that the cellphone dataset is representative of Singapore it is possible to compare the distribution of the home locations identified by our clustering procedure with official census data from 2010~\cite{singapore2010census} (See Figure S10 and Figure S11 in SI). Singapore is divided into 55 urban planning areas\cite{wikipedia} and we compare the number of home locations identified in each of them with the 2010 census data\cite{singapore2010census}. With a correlation coefficient of 0.96, the two spatial distributions are highly linearly correlated as shown in Figure S10 in the supplementary materials. Furthermore, we note that penetration of mobile phones (number of active mobile phone subscriptions compared to the total population) in Singapore was above 140\% at the time of our study~\cite{mobilepenetration}, one of the highest rates in the world. This way, we expect that almost all of the population has a mobile phone and many people have more than subscription. As the flight range of mosquitoes is limited to a few hundred meters~\cite{stoddard2009role}, it is generally assumed that humans carry the dengue virus to previously dengue-free areas and infect local mosquitoes. For this reason in our model mosquitoes don't travel among different cells. Therefore, in the computational implementation each day is represented by two steps: daytime, during which population stay at work, and nightime during which poluation stay at home.

\subsection*{Mobility Models}

We use four different models to estimate mobility of people and assign home and work locations to our agents. The first one is the mobility model defined according to the real estimation of mobility patterns from CDR data as described above. The second mobility model is a model in which for each agent we take the home location from the mobile phone data while the work location is assigned randomly (according to a uniform distribution) among the $2598$ cells. The third mobility model is defined in the following way: for each agent we choose a random home cell of the grid, while the work location is choose with a distance ($d$) that  follows a truncated Levy distribution~\cite{gonzalez2008understanding}  as distribution of the mobility patterns, such as $P(d) \sim (d+d_0)^{-\beta}\exp(-d/k)$, where $P(d)$ is the probability to have of distance $d$ between home and work location, $d_0 (m) = 100$, $\beta=2$ and $k(m)=1500$. The fourth mobility pattern has been generated according to the radiation law of human mobility~\cite{simini2012universal}.  According to this we generated a mobility pattern considering the following: i) we assigned to each cell a number of inhabitants randomly distributed (normal distribution) according to the census data. ii) for each cell we consider that the percentage of commuters is the $80\%$ while the other $20\%$ work and live in the same cell. iii) For all the other inhabitants we computed that distance between home and work location based the radiation laws  that reads $\langle T_{ij}\rangle = T_i \frac{m_i n_j} {(m_i+s_{ij})(m_i + n_{j} + s_{ij})}$, where $T_i$ is the total number of commuters from county $i$, $m_i$ and $n_j$ are the population in county $i$ and $j$ respectively, and $s_{ij}$ is the total population in the circle centered at $i$ and touching $j$ excluding the source and the destination population. The displacement of the agents for the different mobility models are reported in Figure S2 to Figure S5 in the supplementary materials. The generated mobility models show that the radiation model the model generated with the mobile phone are the most similar while there is almost no correlation with the random mobility model as shown in Figure S1 in Supplementary materials. 

\subsection*{Epidemiological Data}
Information about the weekly number of reported Dengue cases in Singapore was collected from the official Singapore's government's one-stop portal\cite{weekly}.  In the 2013 dengue outbreak in Singapore, a significant rise in the number of dengue fever cases was reported in Singapore and caused 8 victims and a total of 22318 cases. In the week of 16–22 June 2013, there was a record of 842 dengue cases in Singapore in a single week. This figure was far beyond the highest number of cases per week in the years 2010, 2011 and 2012. The number of weekly dengue fever cases has exceeded the epidemic threshold of 237. Similarly high number of cases were reported over the course of 2014, with the maximum number of weekly reported cases having a peak of 891. In the following years, the number of dengue cases were significantly lower due to increased efforts to control the mosquito population. We show the total number of dengue cases during 2013 and 2014 in~\figurename~\ref{fig:tempCases}(a). For the spatial analysis of Dengue outbreaks, we use a dataset that is a collection of data from the NEA. Data was collected twice a week since May 2013 (except for a gap in October 2013) from SGCharts Charting Singapore's Data\cite{SGC}. The data provide information of the number of dengue cases in local spatial clusters that were established dynamically based on the location of recent cases. Spatial clusters are typically a few hundred meters in size, encompassing multiple city blocks. This allows us to have a good representation of the spatial spread of dengue, while still protecting to privacy of people affected. We display the cumulative spatial distribution of dengue cases in~\figurename~\ref{fig:spatCase}(a).

\subsection*{Climate Data}
We collected data about climate conditions in Singapore during years 2013 and 2014, during which two outbreaks during the respective summers took place. 
In~\figurename~\ref{fig:tempCases}(a) we show the number of dengue cases during the epidemiological weeks in 2013 and 2014 comparing it with the average temperature. The impact of daily temperature fluctuations on dengue virus transmission by the \emph{A.aegypti} mosquitoes have been extensively studied and the results indicate that the weekly mean temperature is statistically significant relative to the increases in dengue incidence in Singapore and signifies the hazardous impacts of climatic factors on the increase in intensity and magnitude of dengue cases~\cite{hii2009climate}. This reflection can be observed in the outbreaks of 2013 and 2014 where the comparison between reported cases and temperature has been reported in~\figurename~\ref{fig:tempCases}(a).  Weather data including Mean temperature (MeanT, $^{\circ}C$), Minimum temperature (MinT, $^{\circ}C$), Maximum temperature (MaxT, $^{\circ}C$), Rainfall (Rain, mm), Relative humidity (RH, $\%$) and Wind speed (WindS, m/s) were obtained from the National Environment Agency, Singapore (NEA)\cite{rainy}.

\subsection*{Epidemiological Model}
The epidemiological model can be described schematically as shown in~\figurename~\ref{fig:tempCases}(b). Motivated by research that shows that mosquitoes have a very limited flight range and infection is carried by human movement~\cite{stoddard2009role,barmak2011dengue,de2011modeling,stoddard2013house}, we assume mosquitoes to have a fixed location, i.e.~there is no interaction between mosquito populations in distinct grid cells. For this reason, humans are treated as distinct agents, while the values for mosquitoes are aggregated at the cell level. The transitions on the proposed epidemiological model depend on temperature dependent parameters as reported in Table~\ref{table:tempPar} and described in the Supplementary materials (see also Figure S6 and Figure S7 in the supplementary materials). The constant parameters are described in Table~\ref{table:constPar}. 

\begin{table}[h!]
\centering
\caption{Temperature-dependent parameters.} 
\centering 
\begin{tabular}{l   l c } 
\hline\hline 
Notation & Description & Reference \\ [0.5ex] 
\hline 
$\dot{\epsilon_A^v} = \epsilon_A^v (T)$  & transition rate from aquatic to adult mosquito life-stages & \cite{yang2009assessing} \\
$\dot{\mu_A^v} = \mu_A^v (T)$  & mortality rate of aquatic mosquito life-stages & \cite{yang2009assessing} \\
$\dot{\mu_V^v} = \mu_V^v (T)$  & mortality rate of adult mosquito life-stage & \cite{yang2009assessing} \\
$\dot{\theta_V^v} = \theta_V^v (T)$  & intrinsic oviposition rate of adult mosquito life-stage & \cite{yang2009assessing} \\
$\dot{\gamma_V^v} = \gamma_V^v (T)$  & extrinsic incubation period of adult mosquito life-stage & \cite{yang2009assessing} \\
$\dot{\phi}^{h \rightarrow v} = \phi^{h \rightarrow v} (T)$  & human-to-vector probability of transmission per infectious bite & \cite{lambrechts2011impact} \\
$\dot{\phi}^{v \rightarrow h} = \phi^{v \rightarrow h} (T)$  & vector-to-human probability of transmission per infectious bite & \cite{lambrechts2011impact} \\[1ex] 
\hline 
\end{tabular}
\label{table:tempPar} 
\end{table}
\begin{table}[h!]
	\centering
	\caption{Constant parameters.}
	\begin{tabular}{llrc}
		\hline\hline
		Notation & Description & Value & Reference \\ \hline 
		$\gamma^h$ & transition rate from exposed (E) to infected (I) for humans & $0.5\,\mathrm{days}^{-1}$ & \cite{WesolowskiPNAS2015,lourencco20142012} \\
		$\sigma^h$ & recovery rate, i.e.~transition rate from infected (I) to recovered (R) for humans & $0.25\,\mathrm{days}^{-1}$ & \cite{WesolowskiPNAS2015,lourencco20142012} \\
		$c$ & mosquite eggs hatching to larvae & $1$ & \cite{yang2009assessing} \\
		$f$ & female mosquitoes hatched from all eggs & $1$ & \cite{yang2009assessing} \\
	\end{tabular}
	\label{table:constPar}
\end{table}

\subsubsection*{Humans}
In the stochastic framework, we represent each human as an agent $i$, who at each timestep $t$ can be described by a pair $(N,c)_{t,i}$, where $N = S,E,I,R$ is the epidemic state (susceptible, exposed, infected and recovered, respectively), and $c$ denotes the grid cell where the agent resides. In our mobility models, $c$ alternates between a set home and work location, either inferred from the mobile phone usage data in the realistic scenario or generated randomly. We denote by $S_{t,c}$, $E_{t,c}$, $I_{t,c}$ and $R_{t,c}$ the total number of susceptible, exposed, infected and recovered humans in cell $c$ for timestep $t$. We further use $N_{t,c} \equiv S_{t,c} + E_{t,c} + I_{t,c} + R_{t,c}$ for the total number of humans. We assume each agent to be susceptible to the virus initially ($S$). Upon challenge with infectious mosquito bites ($\lambda^{v \rightarrow h}$), individuals enter the incubation phase ($E$) with mean duration of days $1/\gamma^h$, later becoming infectious ($I$) for days $1/\sigma^h$ and finally recovering ($R$) with life-long immunity. 

In each timestep, transition of the states can occur with probabilities $\lambda^{v\rightarrow h}_{t,c} / 2$, $\gamma^h / 2$ and $\sigma^h / 2$ for the case of the $S\rightarrow E$, $E\rightarrow I$ and $I\rightarrow R$ transitions respectively (as each reaction timestep takes half day, we obtain the transition probabilities by halving the daily transition rates). We evaluate the transitions individually for each human agent as a Bernoulli-process, and update the state accordingly. While the $\gamma^h=0.5 \quad days^{-1}$ and $\sigma^h= 0.25 \quad days^{-1}$ rates are constants\cite{lourencco20142012, WesolowskiPNAS2015}, the $\lambda^{v\rightarrow h}_{t,c}$ rate is related to the mosquito population of the grid cell where the human agent is currently residing:
\begin{equation}
	\lambda^{v\rightarrow h}_{t,c} = a \dot{\phi}^{v \rightarrow h} \frac{I^v_{t,c}}{N_{t,c}} = a \dot{\phi}^{v \rightarrow h} \frac{V_{t,c}}{N_{t,c}} \rho^I_{t,c}  \propto V \rho^I
\end{equation}
where $a$ is the biting rate (i.e.~how many humans a mosquito bites on average per day), $\dot{\phi}^{v \rightarrow h}$ is the disease transmission rate per bite, while $I^v_{t,c}$ is the total number of infected mosquitos in cell $c$ at time $t$ (i.e.~$a \frac{I^v_{t,c}}{N_{t,c}}$ gives the probability that an infected mosquito bites the given human agent during this timestep), while $V_{t,c}$ is the total number of  mosquitos in cell $c$ and $\rho^I_{t,c} = I_{t,c} / N_{t,c}$ represents the fraction of infected individuals in that cell. 
The change in compartments of human agents is then expressed by the following equations:
\begin{align}
	t^{S\rightarrow E}_{t,c} = BD(S_{t,c},\lambda^{v\rightarrow h}_{t,c} / 2) \\
	t^{E\rightarrow I}_{t,c} = BD(E_{t,c},\gamma^h / 2) \\
	t^{I\rightarrow R}_{t,c} = BD(I_{t,c},\sigma^h / 2) \\
	S_{t+1,c} = S_{t,c} - t^{S\rightarrow E}_{t,c} \\
	E_{t+1,c} = E_{t,c} + t^{S\rightarrow E}_{t,c} - t^{E\rightarrow I}_{t,c} \\
	I_{t+1,c} = I_{t,c} + t^{E\rightarrow I}_{t,c} - t^{I\rightarrow R}_{t,c} \\
	R_{t+1,c} = R_{t,c} + t^{I\rightarrow R}_{t,c}
\end{align}
where $BD(n,p)$ represents a sample taken from a binomial distribution with $n$ samples and $p$ success probability. We note that during the simulation, the $t$ transition numbers are not calculated by sampling a binomial distribution, but by performing an independent trial for each human agent with the appropriate transition probabilities and recording the number of successes. While the resulting $t$ values are equivalent to sampling a binomial distribution directly, performing the individual trials allow us to track the state of each agent individually. This is necessary to update the populations in the next step based on the movement of agents determined by the mobility model used. 

\subsubsection*{Mosquitoes}
We model the vector population in each grid cell stochastically, where mosquitos have two pertinent life-stages: aquatic (eggs, larvae and pupae, $A$) and adult females ($V$)~\cite{yang2009assessing}. We keep track of the number of mosquitoes for each grid cell and calculate the transmission between the classes stochastically based on the rates calculated from the parameters of the model, some of them being dependent on the temperature. For this, we denote the total number of mosquitoes in each class by $A_{t,c}$ and $V_{t,c}$ respectively for timestep $t$ and cell $c$. We then calculate the changes in mosquito numbers of each mosquito class in each cell according to the following rules.
\begin{align}
    d^A &= BD( A_{t,c}, \dot{\mu}^v_A/2 ) \\
    t^{A\rightarrow V} &= BD( A_{t,c} - d^A, \dot{\epsilon}^v_A/2 ) \\
    d^V &= B( V_{t,c}, \dot{\mu}^v_V /2) \\
    t^{V\rightarrow A} &= PD\left[ cf \dot{\theta}^v_A/2 \left(1 - \frac{A}{K_{t,c}} \right) V \right] 
  \label{eq:mosqchanges1}
\end{align}
and then update the mosquito populations accordingly
\begin{align}
    A_{t+1,c} &= A_{t,c} - d^A - t^{A\rightarrow V} + t^{V\rightarrow A} \\
    V_{t+1,c} &= V_{t,c} - d^V  + t^{A\rightarrow V}
  \label{eq:mosqchanges2}
\end{align}
Here $PD(x)$ represents a sample taken from a Poisson distribution with a mean of $x$. The coefficients $c$ and $f$ are the fraction of eggs hatching to larvae and the fraction of female mosquitoes hatched from all eggs, respectively. For simplicity and lack of quantifications for the local mosquito population, we assume these to be 1~\cite{yang2009assessing}. Moreover, $\dot{\epsilon}^v_A$ denotes the rate of transition from aquatic to adults, $\dot{\mu}^v_A$ and $\dot{\mu}^v_V$ are the mortality rates for aquatic and adult mosquitoes,   $\dot{\theta}^v_A$ is the intrinsic oviposition rates. 
The logistic term  $\left(1 - \frac{A}{K_{t,c}} \right)$ can be understood as the physical/ecological available capacity to receive eggs, scaled by the carrying capacity term $K_{t,c}$ in each cell. The effective carrying capacity $K_{t,c}$ is defined as:
\begin{equation}
K_{t,c} = x_v \frac{W_c + H_c}{2} 
\end{equation}
where $x_v$ is the average number of mosquitos per human, $W_c$ and $H_c$ are respectively the number people whose works or home location is in the cell $c$. This form assumes that the number of mosquitos in a cell scales with the average number of people found there, i.e.~the mean of the nighttime population (defined by the number of home locations in that cell) and daytime population (defined by the number of work locations). Depending on the efficiency of vector control mechanisms, the number of female Aedes mosquitoes per residence varies greatly between countries. In Puerto Rico, the number of mosquitoes per home appears to be between 5 and 10 per home~\cite{newton1992model}, whereas in Singapore, this number is estimated as slightly greater than 0.2~\cite{ooi2006dengue}. This means that the average number of mosquitoes per human in Singapore should be in the range from $0.004$ to $0.01$.  Note that such incorporation of aquatic mosquitoes in our models assumes that every cell contains some breeding sites, which is necessary to sustain a mosquito population if we do not allow mosquitoes to travel between cells.  

All the aquatic mosquitoes ($A^V_t$) that become adult mosquitoes  at time $t$ are susceptible ($S^V_t$) and they can eventually become exposed ($E^V_t$) if they a bite an infected human and they become infected ($I^V_t$) after an incubation time as shown in~\figurename~\ref{fig:tempCases}(b). Both the aquatic and the adult mosquitoes can die with given probabilities ($\mu_A$ and $\mu_V$ respectively). Similarly to the human epidemiological models, the equations describing the vector dynamics are:
\begin{align}
	t^{S^V\rightarrow E^V}_{t,c} = BD(S^V_{t,c},\lambda^{h\rightarrow v}_{t,c} / 2) \\
	t^{E^V\rightarrow I^V}_{t,c} = BD(E^V_{t,c},  \dot{\gamma}^v / 2) \\
	S^V_{t+1,c} = S^V_{t,c} - t^{S^V\rightarrow E^V}_{t,c} \\
	E^V_{t+1,c} = E^V_{t,c} + t^{S^V\rightarrow E^V}_{t,c} - t^{E^V\rightarrow I^V}_{t,c} \\
	I^V_{t+1,c} = I^V_{t,c} + t^{E^V\rightarrow I^V}_{t,c} \\
\end{align}
where the transition rate human-to-vector $\lambda^{h\rightarrow v}_{t,c}$  is defined as\cite{WesolowskiPNAS2015}:
\begin{equation}
	\lambda^{h\rightarrow v}_{t,c} = a \dot{\phi}^{h \rightarrow v} S^V_{t,c}   \frac{I^v_{t,c}}{N_{t,c}}.
\end{equation}
These transitions are function on two temperature dependent parameters such as $\dot{\gamma}^v$ and $\dot{\phi}^{h \rightarrow v}$.

\subsubsection*{Summary}
Using these equations, running the model means iterating the following two steps: 1) Evaluate change of states for every human using individual Bernoulli-trials, and the change in mosquito populations in each cell using Eqs.~(\ref{eq:mosqchanges1}) and~(\ref{eq:mosqchanges2}); 2) Update the locations of human agents based on the mobility model and recalculate the number of humans of each class in each cell accordingly.
We can characterize the mosquito population dynamics and the epidemics based on the ODE representation of the previous model (see SI for the corresponding equations). Using these, we can derive the basic offspring number ($Q$), that is, the mean number of viable female offspring produced by one female adult during its entire time of survival (and in the absence of any density-dependent regulation) as:
\begin{equation}
Q = \frac{\dot{\epsilon}^v_A}{\dot{\epsilon}^v_A + \dot{\mu}^v_A} \frac{cf \dot{\theta}}{\dot{\mu}^v_V}
\end{equation}
All parameters defining $Q$ are temperature-dependent (see below). For a fixed temperature $T_0$ it is possible to derive expressions for the expected population sizes of each mosquito life-stage modelled. These are used to initialize the system, given the temperature present at the initial timepoint:
\begin{equation}
		A(T_0) = K \left(1 - \frac{1}{Q(T_0)} \right) \\
		V(T_0) = K \left(1 - \frac{1}{Q(T_0)} \right) \frac{\dot{\epsilon}^v_A(T_0)}{\dot{\mu}^v_V(T_0)}
\end{equation}

Including the humans, the expression for dengue's basic reproductive number is defined similarly to previous modeling approaches~\cite{wearing2006ecological,lourencco2013natural} but without human mortality: 
\begin{equation}
\dot{R_0} = \frac{V}{N} \frac{ a^2 \dot{\phi}^{v \rightarrow h}}{\sigma^h  \dot{\mu}^v_V}
\label{eq:repnumber}
\end{equation}

We note that as necessary, our model includes some simplifications. Most importantly, parameter values for mosquito population modeling come from controlled experiments performed in laboratory studies~\cite{yang2009assessing}. Clearly, it seems prohibitably challenging to directly estimate these parameters in the wild, as tracking individual mosquitoes is infeasible; studies can test the applicability of the models by comparing predictions to estimates of observed mosquito population sizes. Furthermore, accurately measuring mosquito populations itself present difficulties in real-world conditions. We note that uncertainties in parameters are inherently linked in our model; e.g.~a shorter mosquito lifespan could be offset by higher bite rate as evident from Eq.~\ref{eq:repnumber}. This way, any calibration process among the parameter values will likely be degenerate. Another main limitation in our dataset is that we have no estimate of any existing immunity to dengue in the population. While dengue has mulitple strains, and partial or full immunity can be acquired after being infected with a specific strain, the picture is quite complex. Similarly to uncertainty of parameters for mosquitoes, uncertainty in the size of susceptible population is linked to any variations in other parameters. For this reason, we do not perform a scaling of the population size, but use the sample obtained from the mobile phone data which covers a large part of Singapore's population. We deal with these issues by using established values and temperature-dependent forms from the literature for most parameters~\cite{WesolowskiPNAS2015,lourencco20142012,yang2009assessing}, while exploring a phase space determined by variations in a small number of parameters, namely the bite rate ($a$) and average number of mosquitoes per human ($x_v$). Finding an ideal combination in for this pair of parameters allows us to calibrate the model for Singapore, while avoiding overfitting.

In summary, as initial conditions for the simulations setting we consider $N = 2,307,230$ agents derived from the mobile phone data and described above. At the beginning of the simulations, i.e. January 1st 2013, we set the $I_{init}$ number of initial infected agents as retrieved from the official Singapore's government's portal~\cite{weekly}. In particular $I_{init} = 242$ infected individuals in 93 different cells of the grid $G$. In order to keep the outbreaks \emph{alive} we ensured that  the number of infected individuals in the systems always $I >=100$ as visible in~\figurename~\ref{fig:simCases2013}.
The number of initial aquatic and adult mosquito have been computed for each values of the parameter $x_v$ from January 1st 2011. For each day from January 1st 2011 to December 31st 2012 we collected the temperature and we simulated the dynamics of aquatic and adult mosquitoes in each cell given the population estimated from the mobile phone and following the Eq~\ref{eq:mosqchanges1}  and Eq~\ref{eq:mosqchanges2}. In this way, for each value of the parameter $x_v$ it has been possible to set a stable number of aquatic and adult mosquitoes the first day of the simulation. 

\bibliography{scibib_DK}

\section*{Acknowledgements}
E.M. would like to thank the HERUS Lab at the École polytechnique fédérale de Lausanne, the Swiss Mobiliar insurance company, the ENAC Exploratory Grant 2018 (Preparatory Funding Scheme) with the project entitled ``Risk evaluation of mosquito-borne disease transmission through urban commutes pathways''  and the``Healthy Cities Towards a One Health agenda for urban space'' from the Habitat Research Center at EPFL  for partially founding this research . The authors would like to thank Prof. Alessandro Vespignani for the helpful suggestions regarding the epidemic model.  The authors thank all sponsors and partners of the MIT Senseable City Laboratory including Allianz, the Amsterdam Institute for  Advanced Metropolitan Solutions, the Fraunhofer Institute, Kuwait-MIT Center for Natural Resources and the Environment,  Singapore-MIT Alliance for Research and Technology (SMART) and all the members of the Consortium. 

\section*{Author contributions statement}

E.M. and D.K. conceived the experiments,  E.M. conducted the experiments, E.M. analysed the results, D.K. analyzed the mobile phone data. E.M. and D.K. wrote the paper. All authors reviewed the manuscript. 

\section*{Additional information}
The authors declare that there are not any competing financial and/or non-financial interests in relation to the work described.

\clearpage

\renewcommand{\thefigure}{S\arabic{figure}}
\renewcommand{\thetable}{S\arabic{table}}
\setcounter{figure}{0}
\setcounter{table}{0}
\section*{Supplementary materials\\ \\ Assessing the interplay between human mobility and mosquito borne diseases in urban environments}
Emanuele Massaro$^*$, Daniel Kondor, Carlo Ratti\\
*emanuele.massaro@epfl.ch
\\
\subsection*{Mobility Models}
\begin{figure}[h!]
	\centering
	\includegraphics[width=0.7\linewidth]{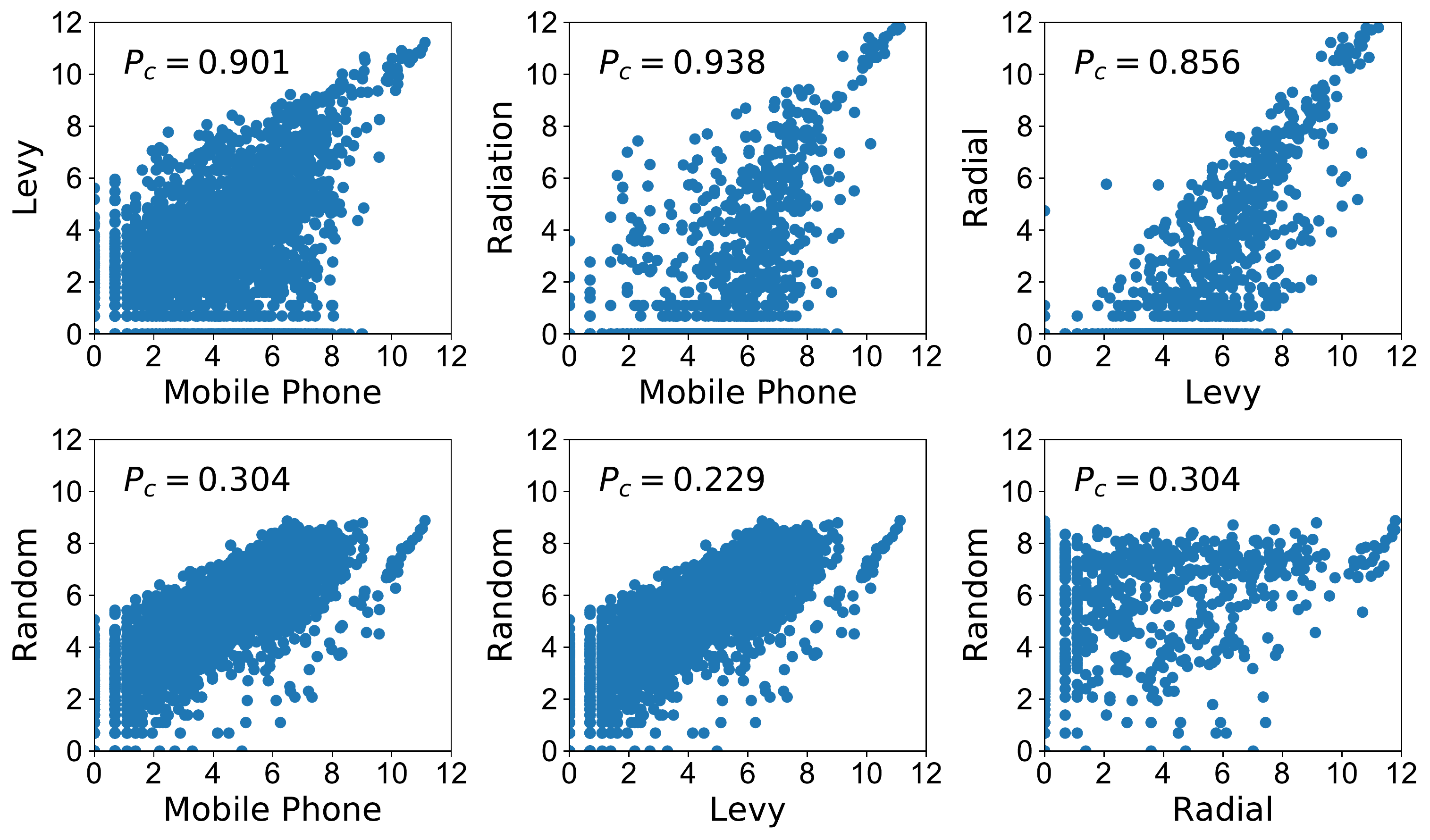}
\caption{Correlation of amount of people travelling between the census areas: each point corresponds to the flow between two census areas.}
\label{fig:flowsCorr}
\end{figure}

\begin{figure}[h!]
    \centering
    \includegraphics[width=1\textwidth]{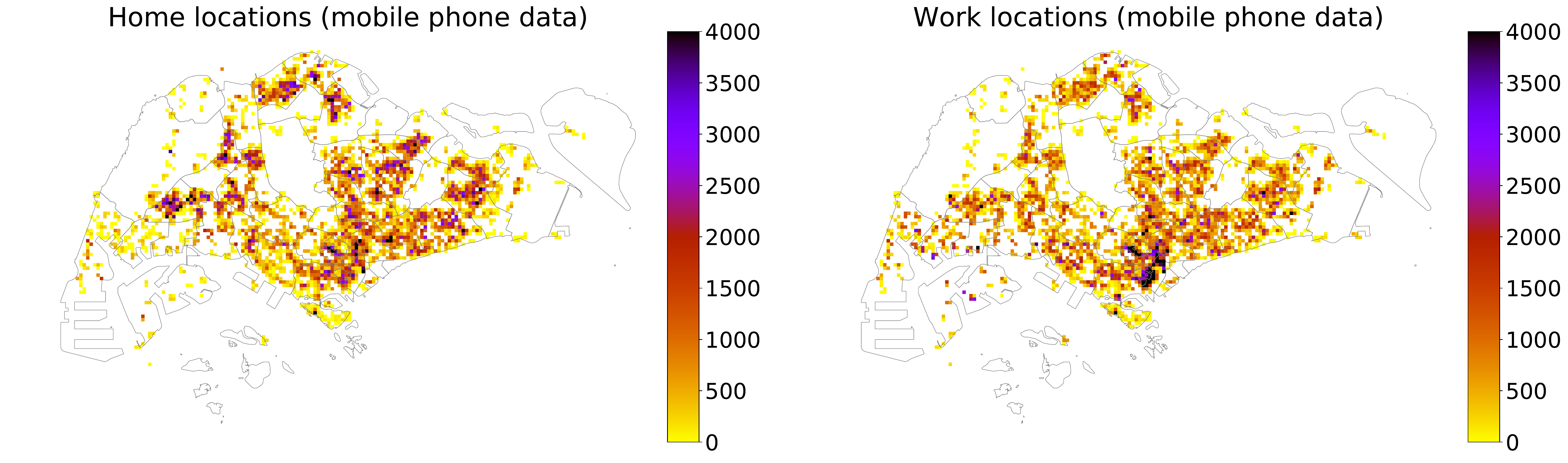}
\caption{\textbf{Home and work locations from mobile phone data}. Count of users in the home and work locations respectively determined from the mobile phone dataset in each cell. The majority of jobs are located in the Central Business District, whereas the home locations are more equally distributed.}
\label{fig:hw}  
\end{figure}

\begin{figure}[h!]
    \centering
    \includegraphics[width=1\textwidth]{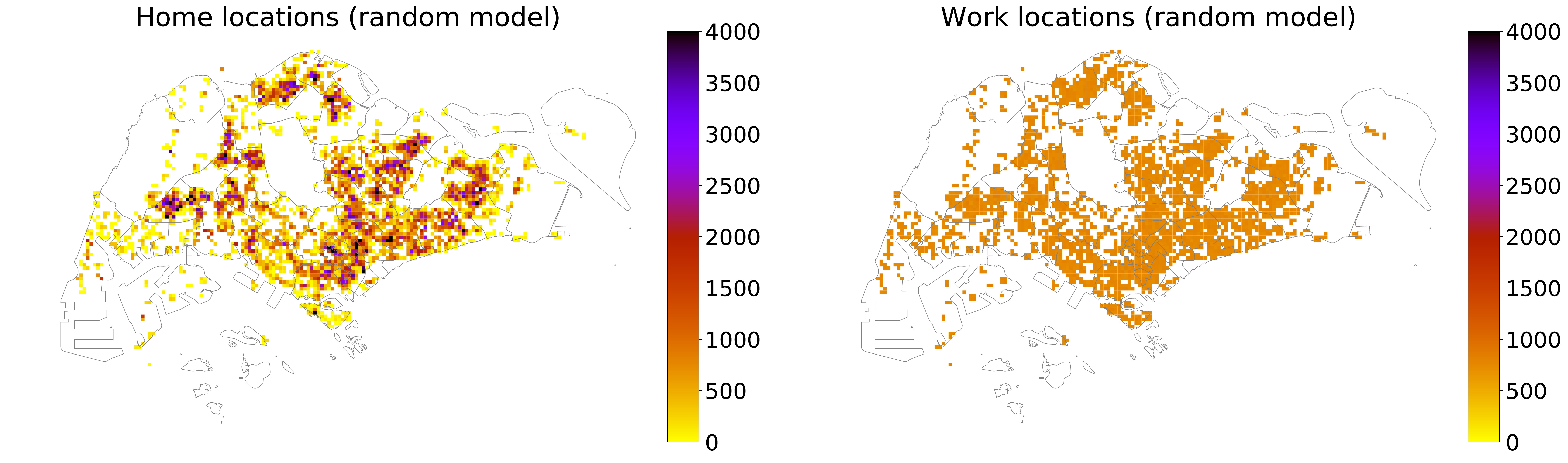}
\caption{\textbf{Home and work locations from the random model}. The home locations are taken from the mobile phone data model while the work locations are randomly assigned.}
\label{fig:hwrand}  
\end{figure}

\begin{figure}[h!]
    \centering
    \includegraphics[width=1\textwidth]{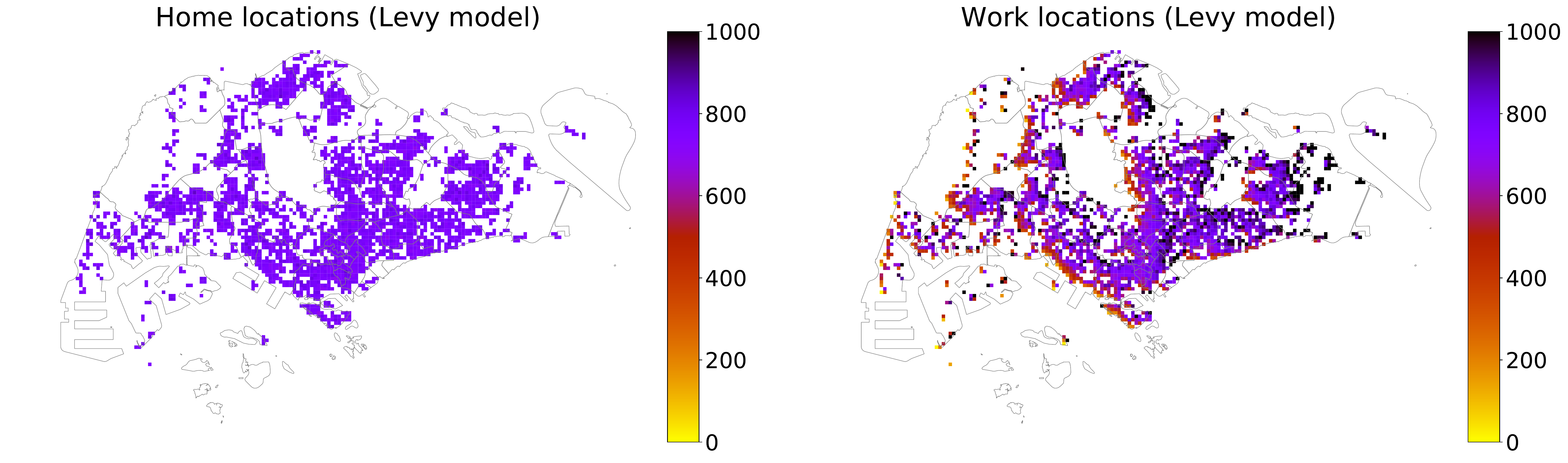}
\caption{\textbf{Home and work locations from the Levy model}. The home locations are randomly assigned while the work locations are given with a distance from a Levy flight distribution.}
\label{fig:hwlevy}  
\end{figure}

\begin{figure}[h!]
    \centering
    \includegraphics[width=1\textwidth]{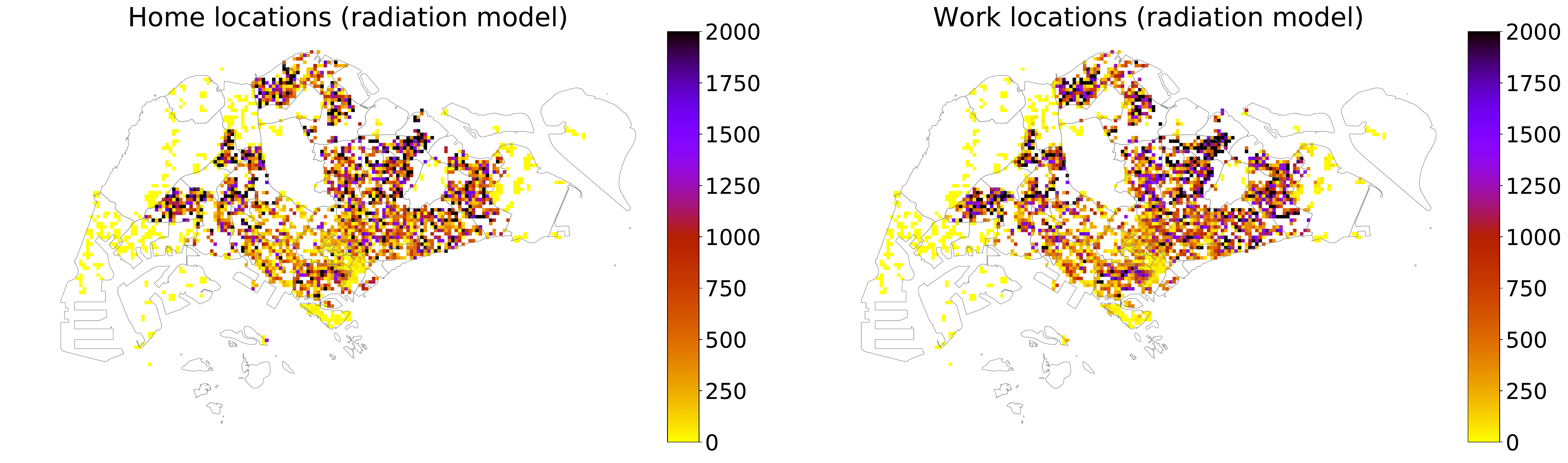}
\caption{\textbf{Home and work locations from the radiation model.}. The home locations are assigned from the census data while the work locations are assigned with a distance following the radiation model.}
\label{fig:hwradial}  
\end{figure}

We report the comparison between the four mobility models used in this research: i) derived from mobile phone data, ii) random, iii) derived from a Levy flight distribution and iv) derived from the radiation model. 
In~\figurename~\ref{fig:flowsCorr} we report the Pearson correlation ($P_c$) coefficient between the 3 models. In the scatterplots each point correspond to the flow (i.e. total number of commuters) from a location $w$ to a location $h$.  mobility models.
\\
 \figurename~\ref{fig:hw}, \figurename~\ref{fig:hwrand}, \figurename~\ref{fig:hwlevy} and   \figurename~\ref{fig:hwradial}  show displacement of the agents in their home and work locations for the different mobility models.

\clearpage
\subsection*{Temperature dependent parameters}
\begin{figure}[h!]
	\centering
	\includegraphics[width=1\linewidth]{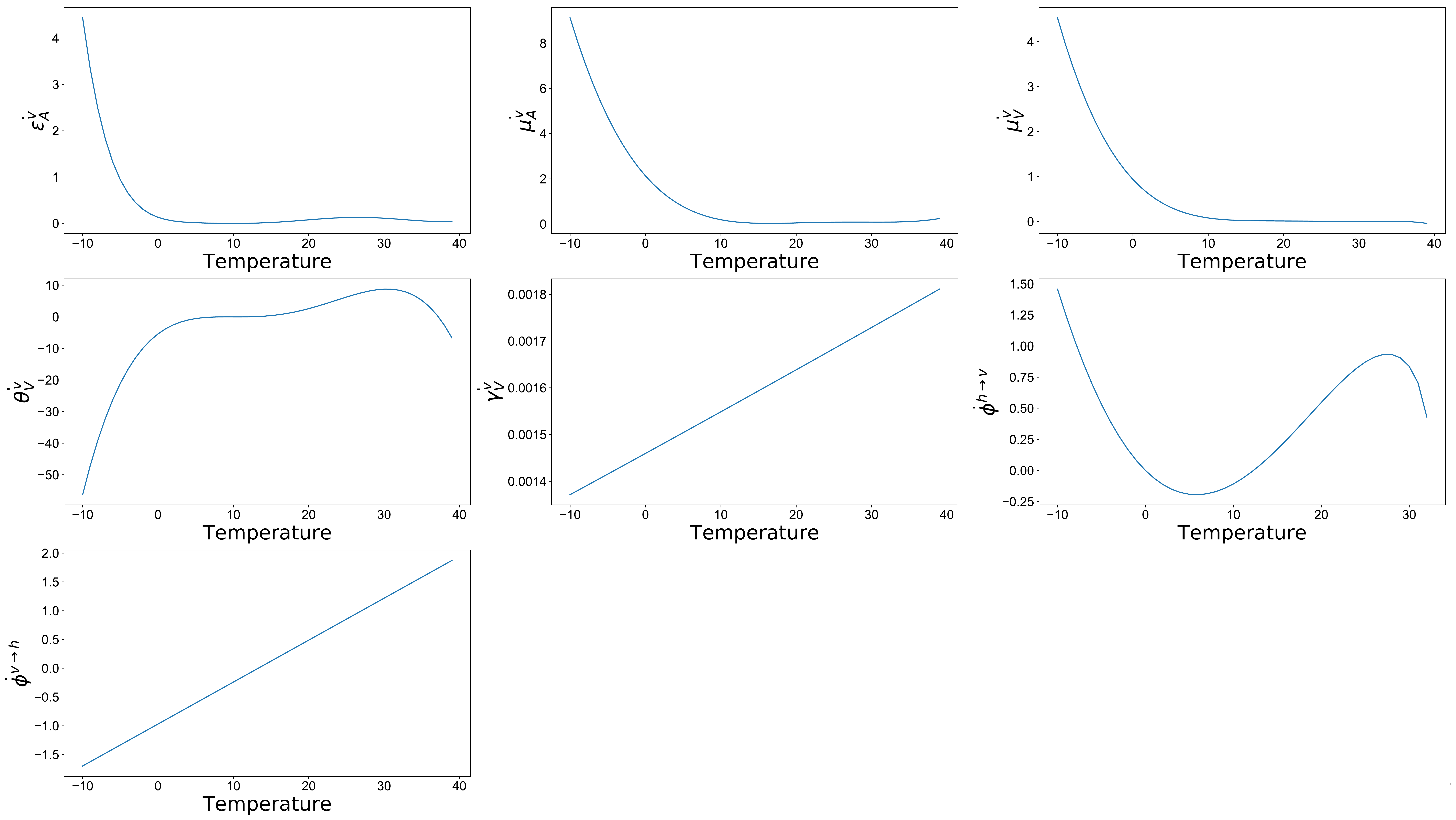}
\caption{The temperature-dependent parameters as function of the temperature.}
\label{fig:partemp}
\end{figure}

\begin{figure}[b!]
	\centering
	\includegraphics[width=1\linewidth]{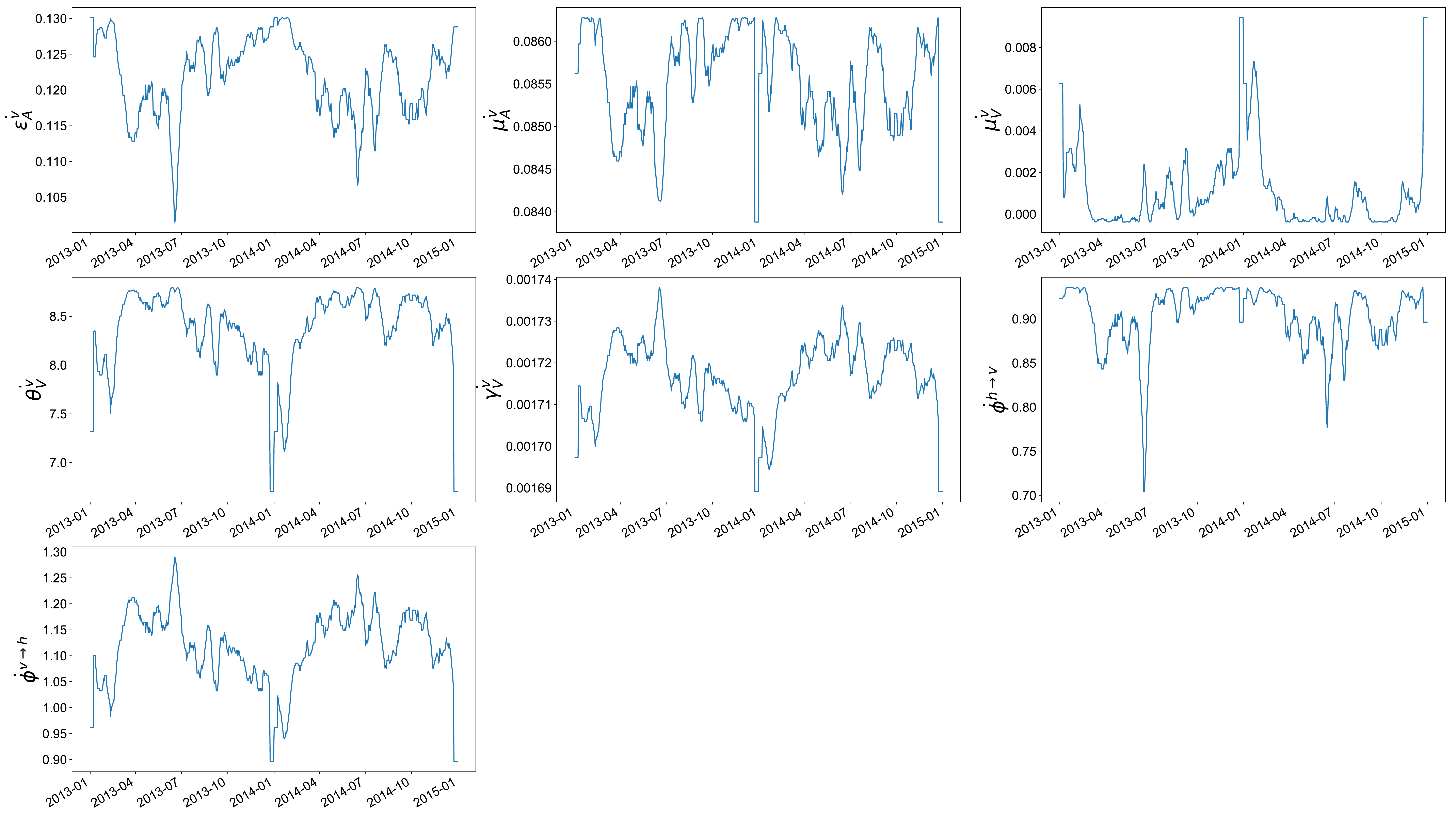}
\caption{The temperature-dependent parameters used in the ento-epidemiological framework for Singapore in 2013-2014.}
\label{fig:tempParameters}
\end{figure}

Most of the parameters (as reported in Table 1  in the main text) used in our methodological approach depend on the temperature. The equations governing those parameters are the following:
\\
\\
$\begin{array}{rcl} 
\epsilon_A^v(T) & = & 0.131 - 0.05723T + 0.01164T^2 - 0.001341T^3 + \\ 
& & + 0.00008723T^4 - 3.017\cdot10^{-6}T^5 + 5.153\cdot10^{-8}T^6 - 3.42\cdot10^{-10}T^7 \\
\mu_A^v(T) & = & 2.13 - 0.3787T + 0.02457T^2  - 6.778\cdot10^{-4}T^3 + 6.794\cdot10^{-6} T^4\\
\mu_V^v(T) & = & RHF * (0.8692 - 0.1599T + 0.01116T^2- 3.408\cdot10^{-4})T^3 + 3.809\cdot10^{-6}T^4) \\
\theta_A^v(T) & = &  -5.4 + 1.8T - 0.2124T^2 + 0.01015T^3 - 1.515\cdot10^{-4}T^4 \\
\gamma_V^v(T) & = & \frac{(3.3589 \cdot 10^{-3} * Tk)/298exp( (1500/R)(1/298-1/Tk))}{1 + exp( ( 6.203\cdot10^{21} )/R * (1/(-2.176 \cdot 10^{30})) - 1/Tk )}
\end{array}
$
\\
\\
where $Tk$ is the degrees in kelvin.
\\
\\
 $\begin{array}{rcl} 
\phi^{h \rightarrow v} (T) & = & 1.004\cdot 10^{-3}  T (T - 12.286) \cdot (32.461-T)^{1/2} \\ 
\phi^{v \rightarrow h} (T) & = & 0.0729T - 0.97.
\end{array}
$
\\
\\
We have also included an adult mortality factor based on relative humidity\cite{WesolowskiPNAS2015}. Temperature and relative humidity are converted to a vapor pressure measure, $VP= 6.11 \cdot 10^{(7.5T/273.3 + T)/ 10}$. This value is converted to a relative  humidity factor RHF) based on the following rules:  If $10<VP<30$, $RHF= 1.2 - 0.2 \cdot VP$, and if $VP \geq 30$, $RHF= 0.5$.
\\
In~\figurename~\ref{fig:partemp} we show the values of the described parameters for temperature between -10\textdegree{}C and 40\textdegree{}C while in~\figurename~\ref{fig:tempParameters} we show the values of the parameters during the years 2013 and 2014.

\clearpage
\subsection*{Structural Similarity Index}
In order to compare and quantify the spatial prediction of the simulations with the real case scenario, we use the \emph{structural similarity index}.
The Structural SIMilarity (SSIM) index is a method originally proposed for measuring the similarity between two images, but is applicable when comparing structural properties of 2-dimensional data, i.e.~the spatial distribution of dengue cases in our case. The SSIM index can be viewed as a quality measure of one of the images being compared, provided the other image is regarded as of perfect quality. It is an improved version of the universal image quality index proposed before~\cite{Wang2004,wang2009mean} and is computed as:
\begin{equation}
SSIM(x,y) = \frac{(2\mu_x \mu_y + c_1)(2\sigma_{xy}+c_2)}{(\mu^2_x\mu^2_y + c_1 )(\sigma^2_x + \sigma^2_y +c_2) }
\end{equation}
where $x$ and $y$ are appropriate-sized windows of the images to compare, where $\mu_x$ and $\mu_y$ are the average of $x$ and $y$, $\sigma^2_x$ and $\sigma^2_y$ are the variances of $x$ and $y$ while $\sigma_{xy}$ is the covariance of $x$ and $y$. The parameters $c_{1}=(k_{1}L)^{2}, c_{2}=(k_{2}L)^{2}$ are two variables to stabilize the division with a weak denominator, where $L$ is the dynamic range of the discrete pixel values. The two additional parameters are $k_{1}=0.01$ and $k_{2}=0.03$ by default. To obtain a similarity metric between two images, the SSIM values are averaged over all possible subsections of the images, defined by sliding windows of size $7\times 7$ pixels.

\begin{figure}[h!]
	\centering
	\includegraphics[width=0.8\linewidth]{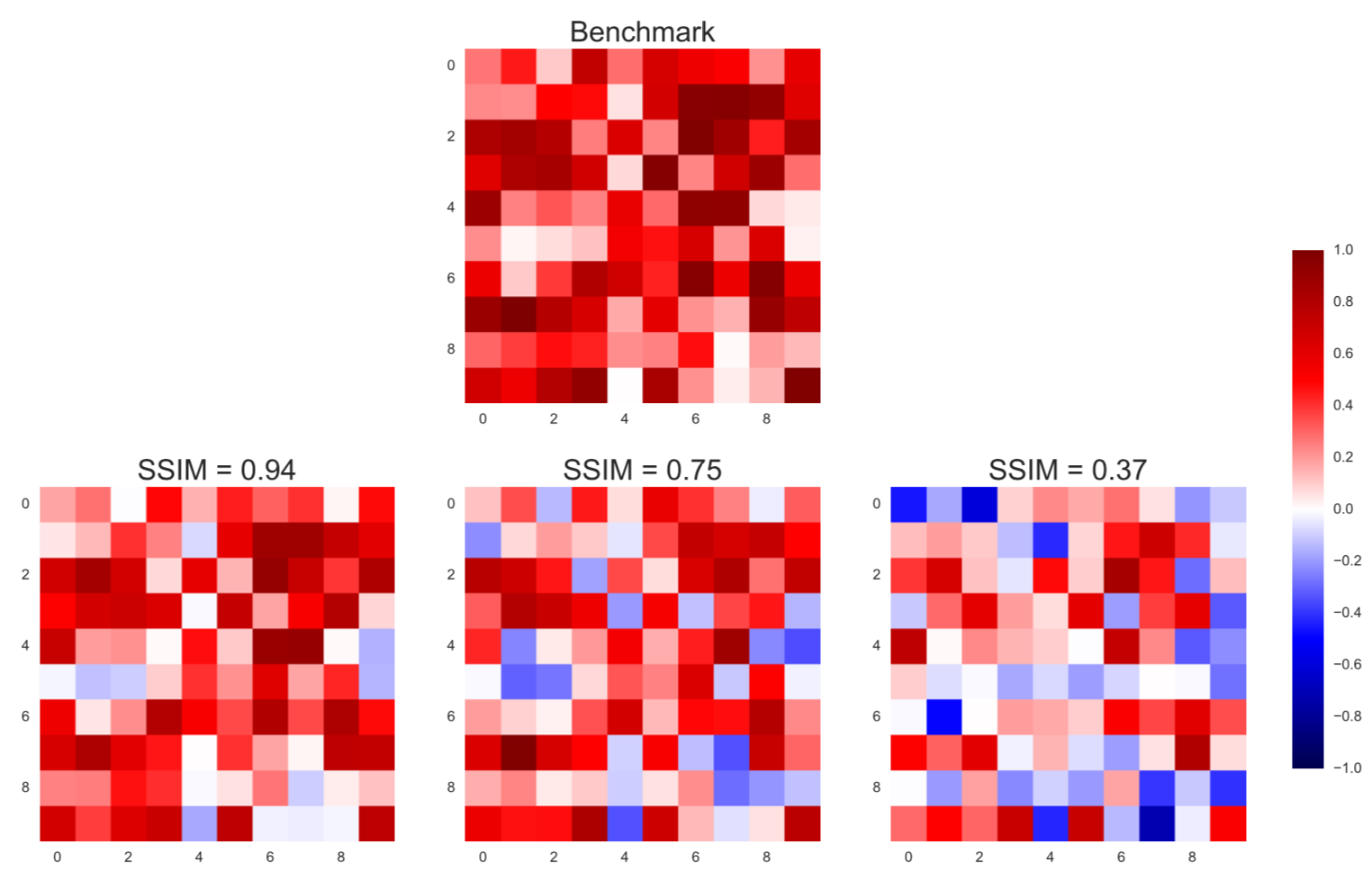}
\caption{Structural similarity index (SSIM) illustration. We generated a $10 \times 10$ grid in which in each cell we assign a random number between $-1$ and $1$. This grid is mathematically described by a matrix $B^{10 \times 10}$ in which each element of the matrix $B_{i,j}$ is a random number between $-1$ and $1$. This matrix represents our benchmark to test the SSIM index. We then generate other three grids (from left to right bottom) starting from the benchmark in which for each $B_{i,j}$ we add or subtract random number between  $0$ and $1$ times $0.25$, $0.5$ and $0.75$ respectively. In this way we are able to compare three different scenarios with the benchmark with different degree of difference from the original one. As we can observe similar images generate greater SSIM if compared with the benchmark.}
\label{fig:SSIMEx}
\end{figure}

The range of the value of the SSIM index is between $0$ and $1$: when two images are nearly identical, their SSIM is close to $1$.
For each epidemiological week in the period we compute the SSIM between the real case and the three simulated scenarios using the the \texttt{Python} function \texttt{structural\_similarity} from the package \texttt{skimage}\footnote{\url{http://scikit-image.org/docs/dev/auto\_examples/transform/plot\_ssim.html}}. An example of SSIM in a toy grid is reported in~\figurename~\ref{fig:SSIMEx}.

\clearpage

\begin{figure}[b!]
    \centering
    \begin{subfigure}[b]{0.5\textwidth}
        \centering
        \includegraphics[width=1\linewidth]{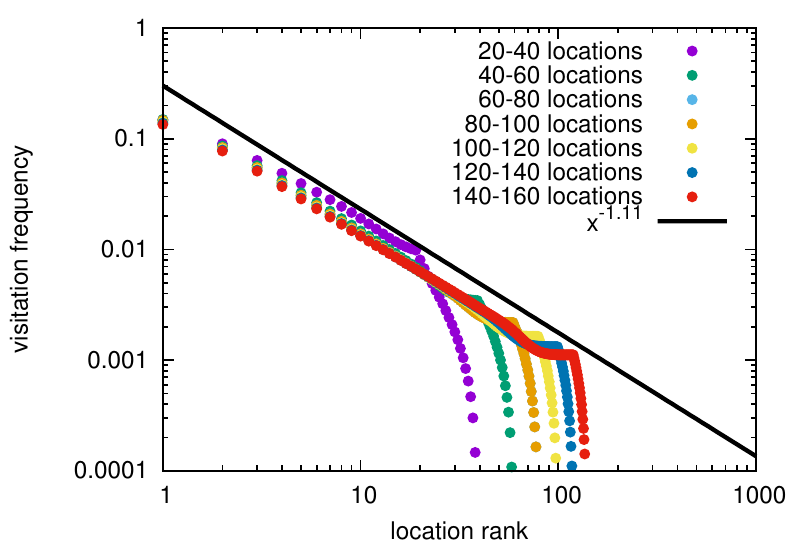}
        \caption{}
    \end{subfigure}%
    ~ 
    \begin{subfigure}[b]{0.5\textwidth}
        \centering
        \includegraphics[width=1\linewidth]{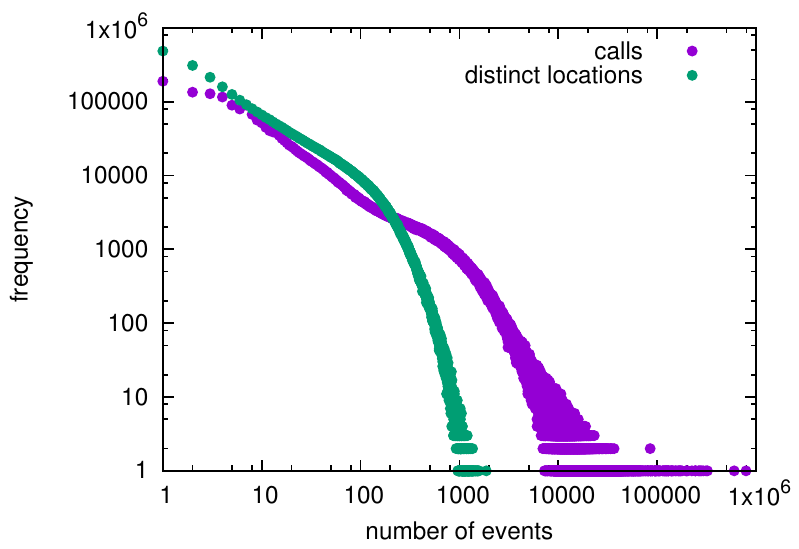}
        \caption{}
   \end{subfigure}%
     \\
    \begin{subfigure}[b]{0.5\textwidth}
        \centering
        \includegraphics[width=1\linewidth]{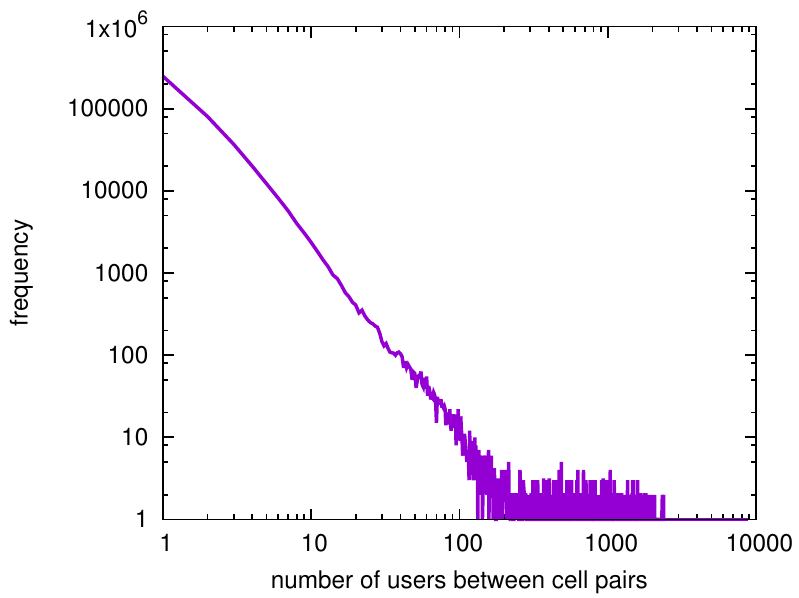}
        \caption{}
        \label{fig:s1c}
    \end{subfigure}%
    \begin{subfigure}[b]{0.5\textwidth}
        \centering
        \includegraphics[width=1\linewidth]{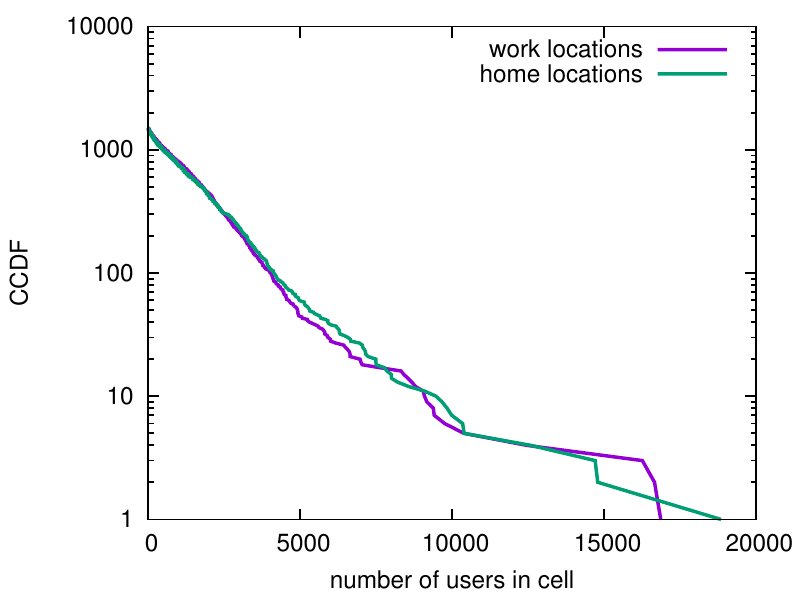}
        \caption{}
         \label{fig:s1d}
    \end{subfigure}
\caption{Statistical analysis of mobile phone data. A) Distribution of number of locations (antennas). B) events (calls / texts) per user and distribution of visitation frequencies of user locations as a function of location ranks (here the locations still refer to antennas). C) Distribution of number of locations (antennas) and (b) events (calls / texts) per user and distribution of visitation frequencies of user locations as a function of location ranks (here the locations still refer to antennas). D) Distribution of the commute matrix elements (i.e. the number of users who commute between any two cells).}
\label{fig:s1}  
\end{figure}

\begin{figure}[t!]
	\centering
	\includegraphics[width=0.5\linewidth]{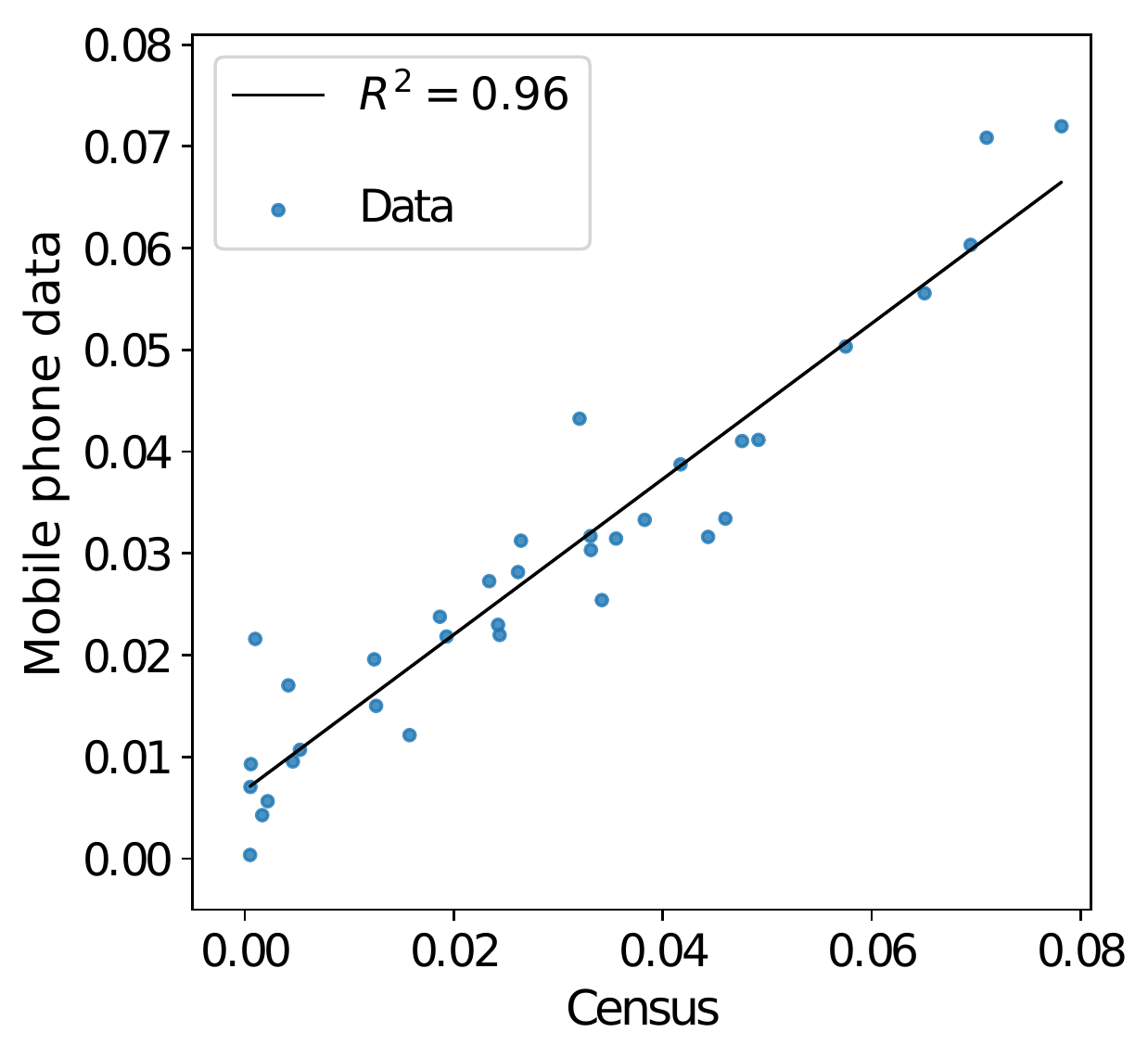}
\caption{\textbf{Fraction of the population in Singapore's districts} according to the 2010 census versus the home locations determined from the mobile phone dataset. With a correlation coefficient of 0.96, the two spatial distributions are highly linearly correlated.}
\label{fig:singcensus}
\end{figure}

\begin{figure}[h!]
	\centering
	\includegraphics[width=\linewidth]{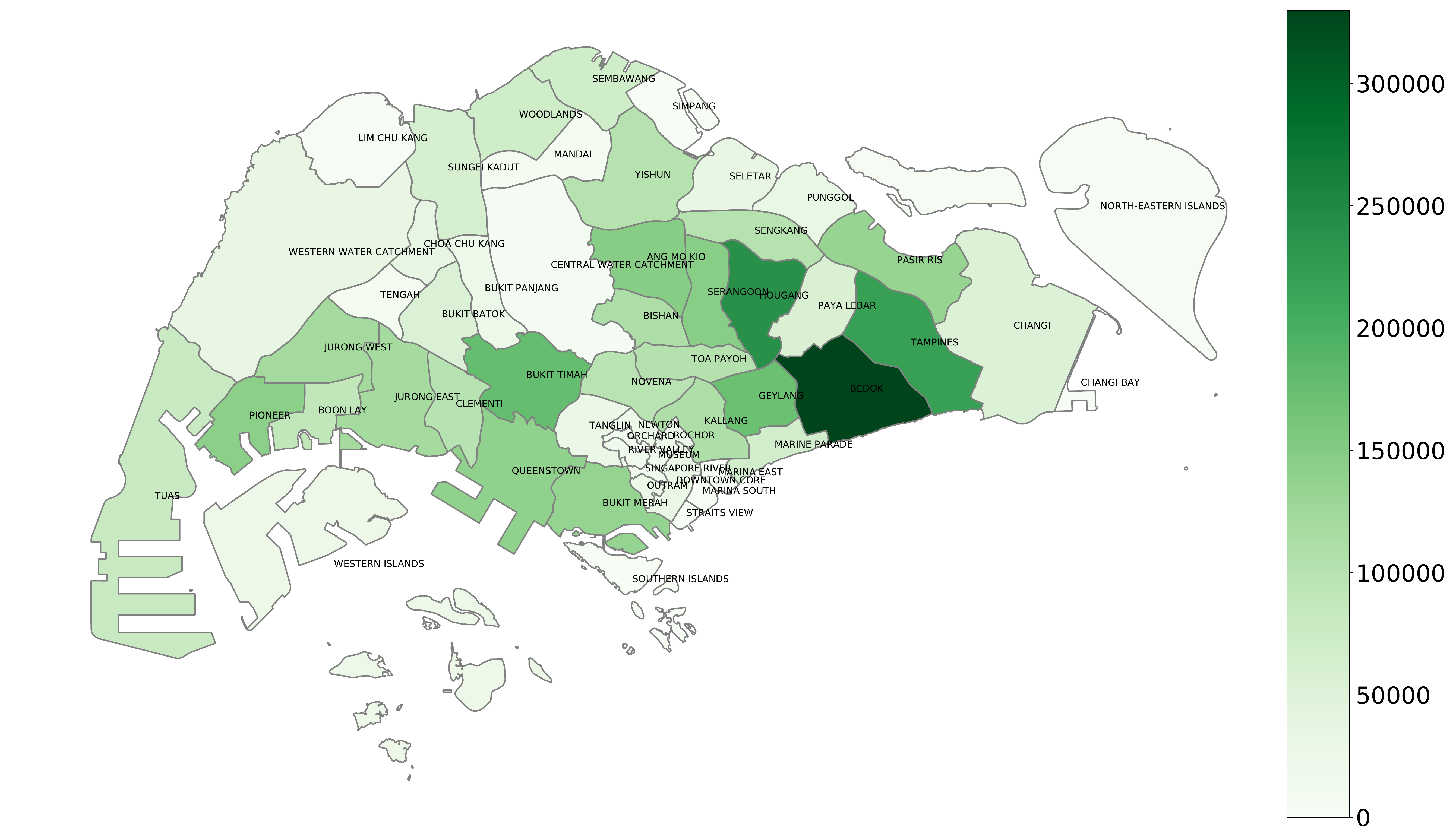}
\caption{Basic demographic characteristics of the Singapore resident population by their registered place of address from the Census of Population 2010. The Singapore resident population comprises Singapore citizens and permanent residents. Of the $3.77$ million Singapore residents as at end-June 2010, about $57\%$ were concentrated in ten planning areas. There were five planning areas with more than $200,000$ Singapore residents. Bedok, Jurong West and Tampines each had more than $250,000$ Singapore residents, with Bedok having the most number at $294,500$ in 2010. The other two planning areas with more than $200,000$ Singapore residents in 2010 were Woodlands ($245,100$) and Hougang ($216,700$). Shapefile data are downloaded from the Singapore open data portal https://data.gov.sg, while population data are downloaded from https://www.worldpop.org/}	
\label{fig:s2}
\end{figure}

\begin{figure}[h!]
	\centering
	\includegraphics[width=\linewidth]{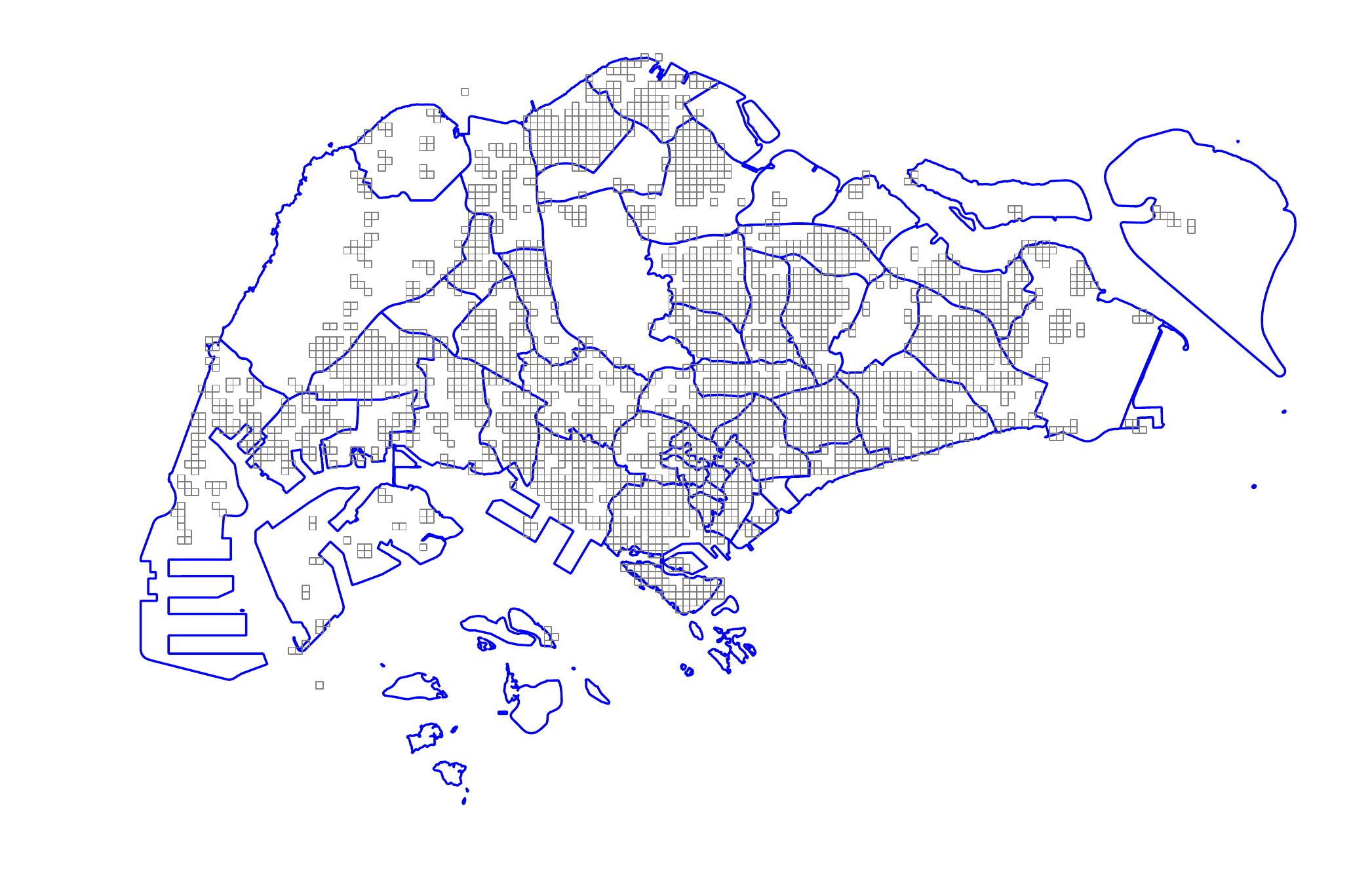}
\caption{The 2598 cells used in this research.}	
\label{fig:grids}
\end{figure}

\begin{figure}[b!]
    \centering
    \begin{subfigure}[b]{0.5\textwidth}
        \centering
        \includegraphics[width=1\linewidth]{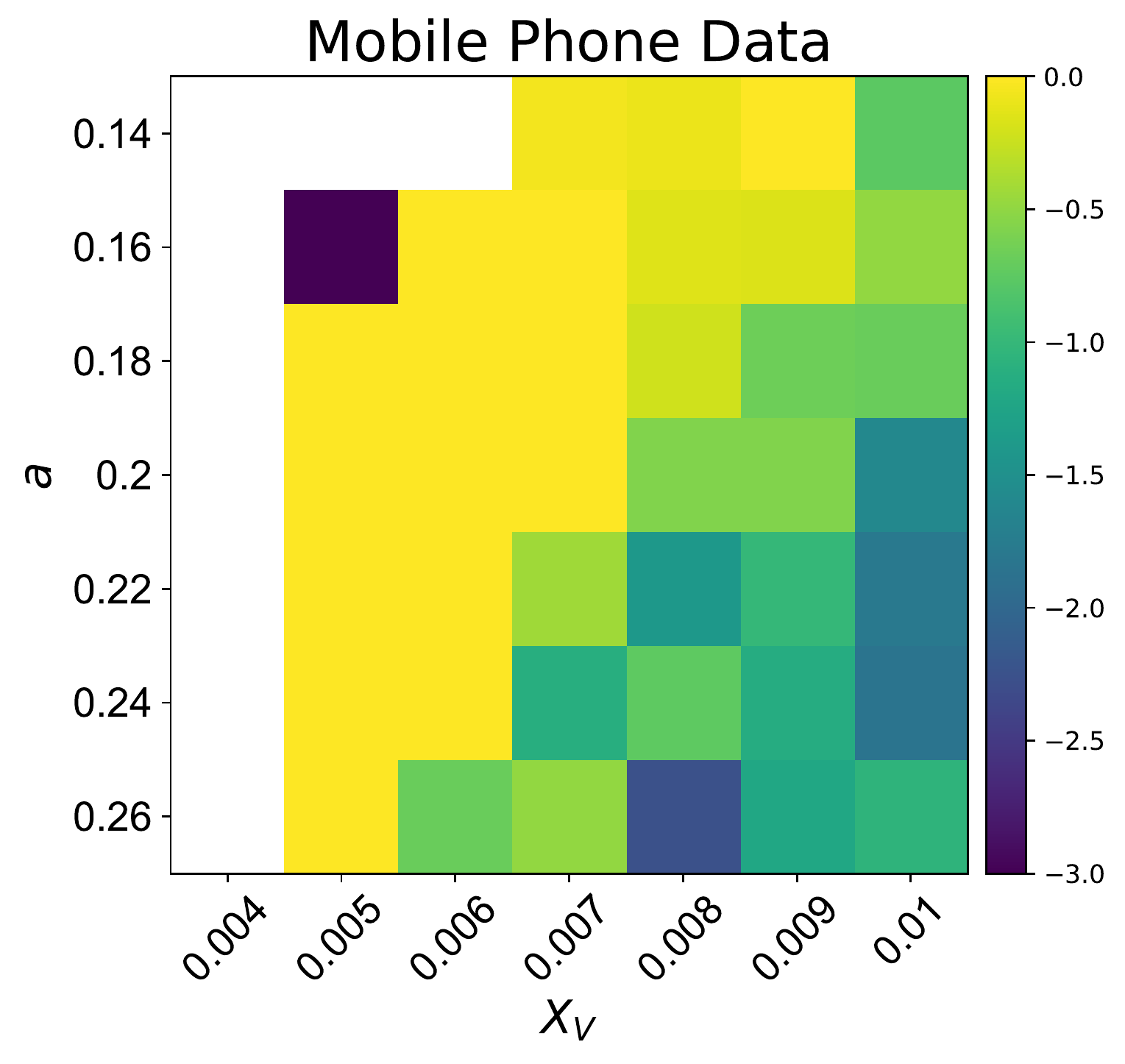}
    \end{subfigure}%
    ~ 
    \begin{subfigure}[b]{0.5\textwidth}
        \centering
        \includegraphics[width=1\linewidth]{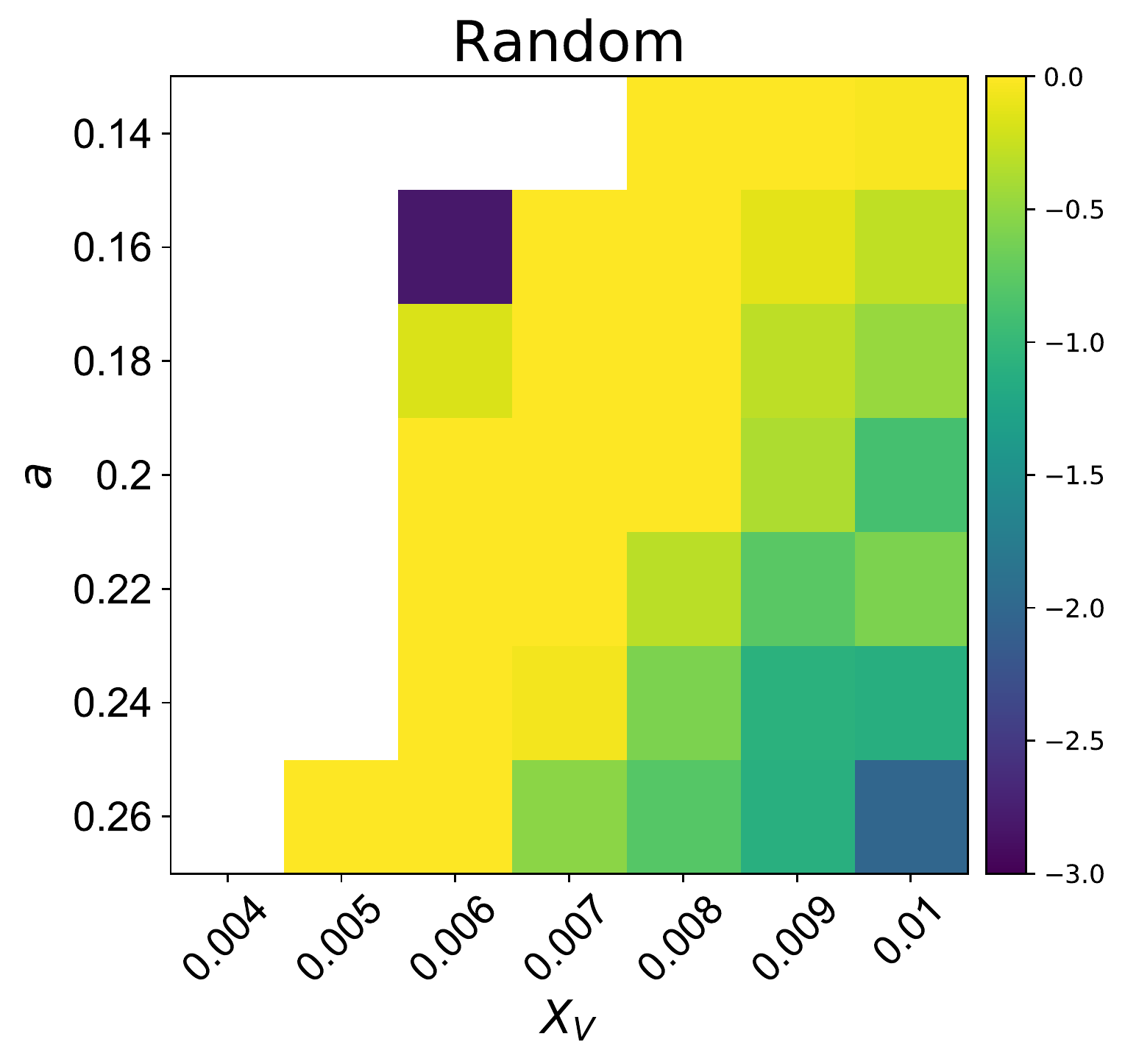}
   \end{subfigure}%
     \\
    \begin{subfigure}[b]{0.5\textwidth}
        \centering
		\includegraphics[width=1\linewidth]{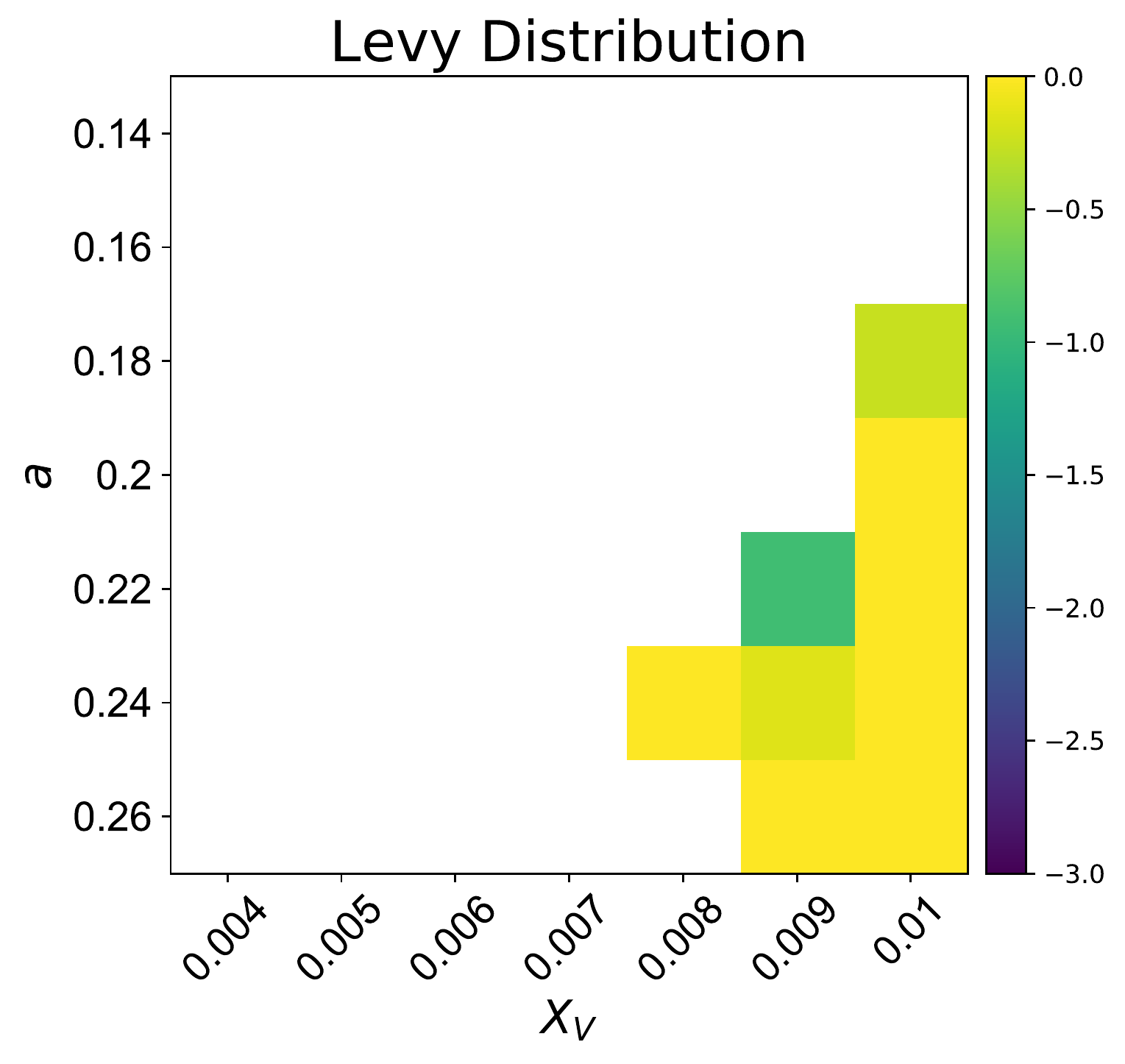}
    \end{subfigure}%
    \begin{subfigure}[b]{0.5\textwidth}
        \centering
        \includegraphics[width=1\linewidth]{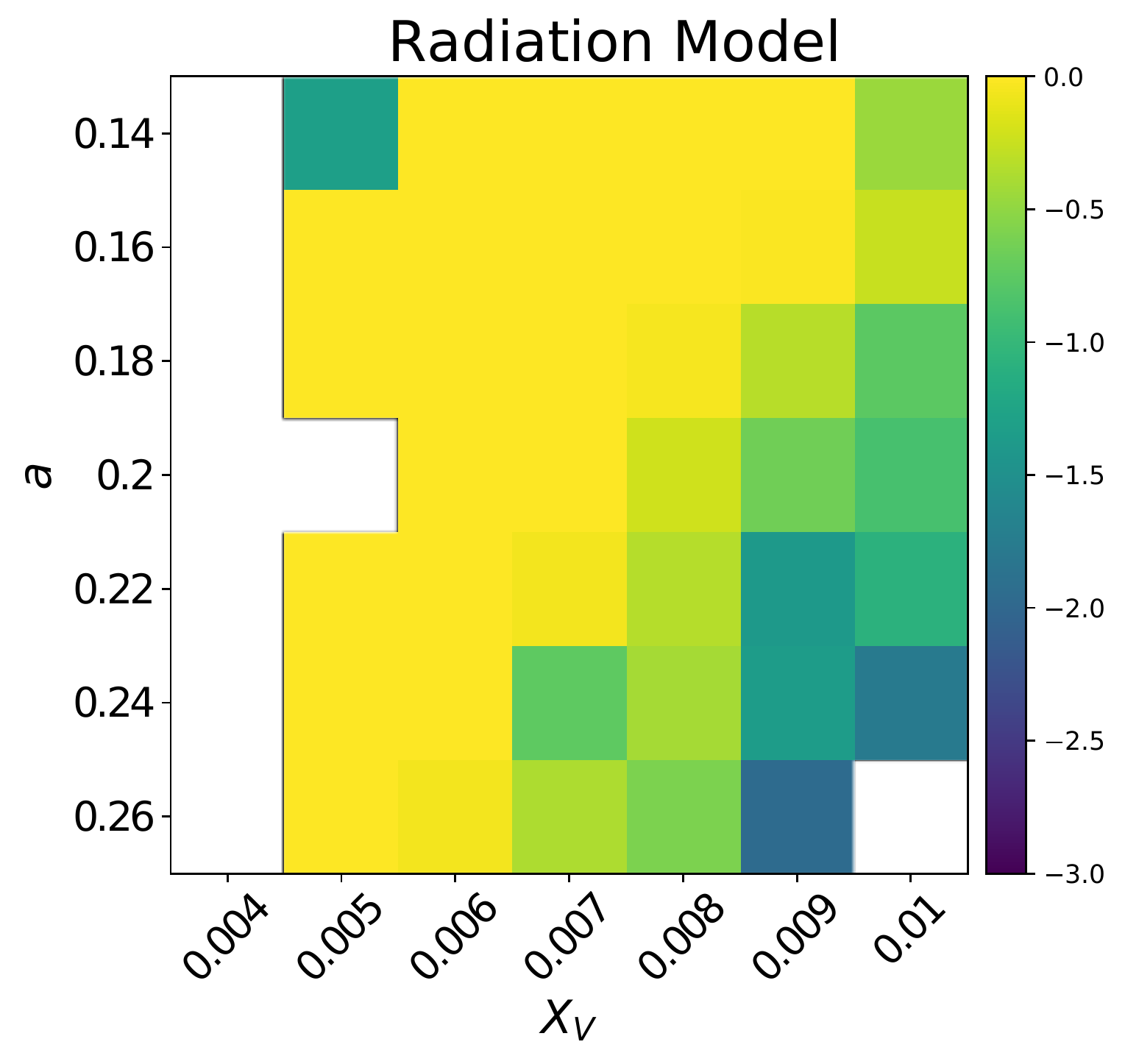}
    \end{subfigure}
\caption{Logarithmic  values of the $R^2$ of the temporal estimation for each couple of parameters in the different mobility models where the white spaces correspond to the negative value of the $R^2$.} 
\label{fig:sIM}  
\end{figure}

\end{document}